\DocumentMetadata{}

\documentclass[authorversion, sigconf]{acmart}

\usepackage{wrapfig}
\usepackage {booktabs}
\usepackage{tikz}

\AtBeginDocument{%
  }

\copyrightyear{2024}
\acmYear{2024}
\setcopyright{rightsretained}
\acmConference[AHs 2024]{The Augmented Humans International Conference}{April 4--6, 2024}{Melbourne, VIC, Australia}
\acmBooktitle{The Augmented Humans International Conference (AHs 2024), April 4--6, 2024, Melbourne, VIC, Australia} 
\acmDOI{10.1145/3652920.3652925} 
\acmISBN{979-8-4007-0980-7/24/04}

\begin{document}

    \title{WhisperMask: A Noise Suppressive Mask-Type Microphone for Whisper Speech}


\author{Hirotaka Hiraki}
\affiliation{%
  \institution{The University of Tokyo}
  \streetaddress{7-3-1, Hongo}
  \city{Bunkyo}
  \state{Tokyo}
  \country{Japan}
  \postcode{43017-6221}
}
\affiliation{%
  \institution{National Institute of Advanced Industrial Science and Technology}
  \streetaddress{6-2-3, Kashiwanoha}
  \city{Kashiwa}
  \state{Chiba}
  \country{Japan}
}
\email{hirotakahiraki@gmail.com}

\author{Shusuke Kanazawa}
\affiliation{%
  \institution{National Institute of Advanced Industrial Science and Technology}
  \streetaddress{6-2-3, Kashiwanoha}
  \city{Kashiwa}
  \state{Chiba}
  \country{Japan}
}
\email{kanazawa-s@aist.go.jp}

\author{Takahiro Miura}
\affiliation{%
  \institution{National Institute of Advanced Industrial Science and Technology}
  \streetaddress{6-2-3, Kashiwanoha}
  \city{Kashiwa}
  \state{Chiba}
  \country{Japan}
}
\email{miura-t@aist.go.jp}

\author{Manabu Yoshida}
\affiliation{%
  \institution{National Institute of Advanced Industrial Science and Technology}
  \streetaddress{6-2-3, Kashiwanoha}
  \city{Kashiwa}
  \state{Chiba}
  \country{Japan}
}
\email{yoshida-manabu@aist.go.jp}

\author{Masaaki Mochimaru}
\affiliation{%
  \institution{National Institute of Advanced Industrial Science and Technology}
  \streetaddress{6-2-3, Kashiwanoha}
  \city{Kashiwa}
  \state{Chiba}
  \country{Japan}
}
\email{m-mochimaru@aist.go.jp}

\author{Jun Rekimoto}
\affiliation{%
  \institution{The University of Tokyo}
  \streetaddress{7-3-1, Hongo}
  \city{Bunkyo}
  \state{Tokyo}
  \country{Japan}
  \postcode{43017-6221}
}
\affiliation{%
  \institution{Sony Computer Science Laboratory}
  \streetaddress{7-3-1, Kyotoxt}
  \city{Bunkyo}
  \state{Tokyo}
  \country{Japan}
  \postcode{43017-6221}
}
\email{rekimoto@acm.org}

\begin{teaserfigure}
  \includegraphics[width=\textwidth]{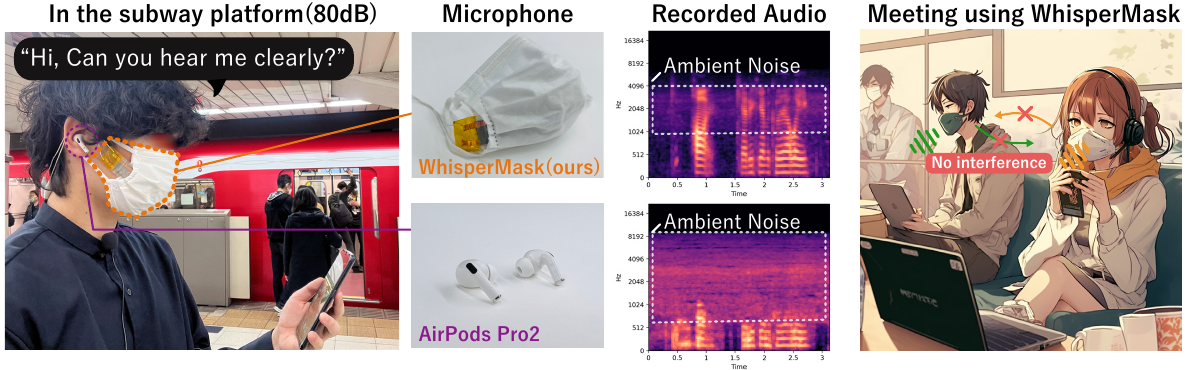}
  \caption{WhisperMask is a wearable mask-type microphone that captures only the user’s voice even in noisy environments, such as in subway stations. The audio from WhisperMask has lower ambient noise(upper middle) than that recorded from other wearable microphones, such as Apple AirPods Pro2. In indoor setting, the WhisperMask did not pick up background noises during voice calls.} 
  \label{fig:teaser}
\end{teaserfigure}


\newcommand{\doublecirc}{{\ooalign{$\bigcirc$\crcr\hss$\circ$\hss}}}

\begin{abstract}



Whispering is a common privacy-preserving technique in voice-based interactions, but its effectiveness is limited in noisy environments. In conventional hardware- and software-based noise reduction approaches, isolating whispered speech from ambient noise and other speech sounds remains a challenge. We thus propose WhisperMask, a mask-type microphone featuring a large diaphragm with low sensitivity, making the wearer's voice significantly louder than the background noise.
We evaluated WhisperMask using three key metrics: signal-to-noise ratio, quality of recorded voices, and speech recognition rate. Across all metrics, WhisperMask consistently outperformed traditional noise-suppressing microphones and software-based solutions. Notably, WhisperMask showed a 30\% higher recognition accuracy for whispered speech recorded in an environment with 80 dB background noise compared with the pin microphone and earbuds.
Furthermore, while a denoiser decreased the whispered speech recognition rate of these two microphones by approximately 20\% at 30-60 dB noise, WhisperMask maintained a high performance even without denoising, surpassing the other microphones' performances by a significant margin.

\begin{figure*}[htbp]
    \centering
    \includegraphics[width=\textwidth]{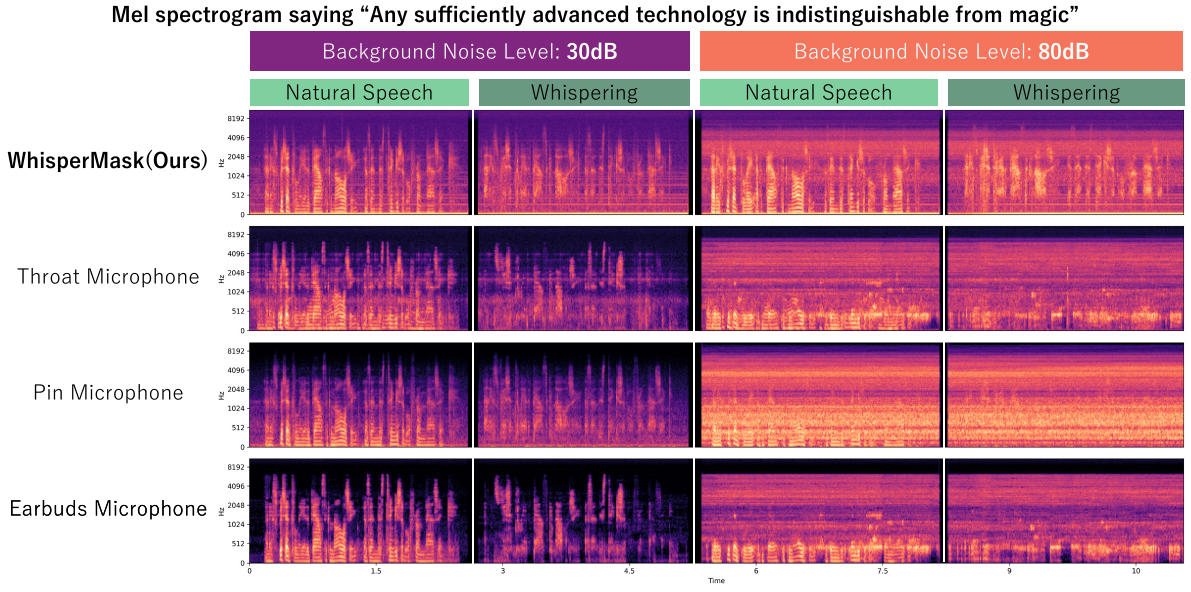}
    \caption{Mel-spectrogram of the speech ”Any sufficiently advanced technology is indistinguishable from magic.” Displayed are two speaking styles: natural speech(left side) and whispered speech(right side) delivered in an environment with varying noise levels (30 dB and 80 dB) and recorded using different microphones. The topmost Mel spectrogram shows that the proposed microphone effectively captures whispers at 80 dB noise level.}
    \label{fig:sample_of_speech_in_whispermask_and_other_microphones}
\end{figure*}

WhisperMask's design renders the wearer's voice as the dominant input and effectively suppresses background noise without relying on signal processing. This device allows for reliable voice interactions, such as phone calls and voice commands, in a wide range of noisy real-world scenarios while preserving user privacy.

\end{abstract}




\maketitle

\section{Introduction}
With the widespread use of smart devices and voice interfaces, on-the-go communication is becoming increasingly common. Microphones are now integrated into various wearable devices, such as AirPods\cite{appleinsider.com}, smart watches\cite{smartwatch_mic1, smartwatch_mic2}, and smart rings\cite{smartring_mic1}. However, using these interfaces in noisy environments poses technical challenges, as the wearer's voice can be interfered with by other speakers' voices and background noise\cite{speech_recognition_survey}.
The existing solutions to achieve a clear voice input in noisy environments include hardware advancements, such as the development of unidirectional microphones\cite{A-century-of-microphones}, throat microphones\cite{Throat-microphone}, and non-audible murmur microphones\cite{nam_microphone}, as well as software-based approaches such as blind source separation (BSS)\cite{bss_review1,bss_survey_sawada_ono_kameoka_kitamura_saruwatari_2019} and speech enhancement\cite{enhancement_review1, enhancement_review2}. However, these solutions have limitations in terms of wearability, contact noise, and the ability to capture whispered speech (Table \ref{tab:type_of_microphone}).
Some products, such as HashMe\cite{Hushme} and Mutalk\cite{mutalk}, aim to capture only the wearer's voice in noisy environments firmly attached around the mouth. However, they limit face-to-face conversations and they can be bulky. Microphones with built-in ventilation fans\cite{LGPuriCare2024} have also been proposed, but they generate significant noise for those nearby.
To address these issues, we propose WhisperMask, a lightweight mask-type microphone with a large, soft diaphragm made of conductive fabric. Unlike the existing mask-type microphones that are firmly attached around the mouth. WhisperMask allows for natural breathing and speaking while effectively suppressing background noise without relying on signal processing. We evaluate WhisperMask using three metrics, namely, signal-to-noise ratio (SNR), audio quality, and speech recognition rate, to demonstrate its superior performance compared with that of conventional noise-suppressive microphones and software-based solutions. Our main contributions are as follows:

		
    
     

\begin{table*}[htbp]
\centering
\caption{Types of microphones based on design principle. WhisperMask is wearable, eliminates background or contact noise, and capable of capturing whispered voices even in noisy places.}
\begin{tabular}{llcccc}\toprule
Microphone & Principle of Noise Reduction & No background noise & Wearable & No contact noise & Whispering \\\midrule
Unidirectional Mic & Directional Vibration & $\bigtriangleup$ & $\checkmark$ & $\checkmark$ & $\checkmark$ \\
Array Mic & Beamforming & $\bigtriangleup$ & & $\checkmark$ & $\checkmark$ \\
Throat Mic & Piezoelectric & $\bigtriangleup$ & $\checkmark$ & & \\
Nam Mic & Contact Microphone & $\bigcirc$ & $\checkmark$ & & \\
Earbuds Mic & Deep Learning & $\bigtriangleup$ & $\checkmark$ & $\checkmark$ & \\
\textbf{WhisperMask} & \textbf{Large Electlet Diaphragm} &$\bigcirc$ &$\checkmark$ &$\checkmark$&$\checkmark$ \\ \bottomrule
\end{tabular}
\label{tab:type_of_microphone}
\end{table*}

\begin{itemize}
\item Development of WhisperMask, a wearable, noise-suppressing microphone that captures whispered speech in environments with up to 80 dB ambient noise.
\item Acoustic characterization, which revealed that WhisperMask provides a 10 dB advantage to the wearer’s voice over any external noise ranging from 200 Hz to 5 kHz.
\item SNR evaluation, which showed that WhisperMask outperforms the existing microphones by 10 dB in environments with 70 dB noise.
\item Audio quality assessment, which demonstrated WhisperMask's superiority over the other microphones and its comparability to noise reduction software.
\item Speech recognition evaluation, which showed that WhisperMask achieved a 30\% higher recognition rate than a noise suppression software for both normal and whispered speech in noisy conditions.
\end{itemize}

In this study, we explore a new microphone diaphragm shape and a mask-type interface that captures only the user's voice without sealing the mouth, enabling reliable voice interactions in noisy real-world scenarios while preserving user comfort and privacy. As shown in Fig 2, the proposed WhisperMask effectively captures both normal speech and whispered speech in high-noise environments, demonstrating its potential for various voice interaction applications.

\section{Related Work}
Voice interaction is one of the most important interaction modalities that people can engage in, and it is used in various applications such as telephone and online calls, voice command input, and interactive operation with smart assistants. In these applications, clear speech input is important; equally important is a stable voice interaction even in environments where other people are talking or where a high noise level exists, such as in subways or construction sites. To achieve clear speech input, many approaches have long been proposed in the field of speech, signal processing, and interaction, both from hardware and software perspectives.

One particular approach is increasing the number of microphones, allowing devices to become more directional and select toward a user's voice \cite{bss_microphone_work_well_with_numbers}; however, such devices are not suitable to become wearable due to their large size. In this section, we present the existing approaches that facilitate voice interaction designed to be wearable and operate in real-time.

\subsection{Wearable noise-suppressing microphones}
In terms of hardware technology, noise suppression is achieved by combining the principles of sound pickup around which various microphones are designed, including unidirectional microphones \cite{A-century-of-microphones}, throat microphones \cite{Throat-microphone}, NAM microphones \cite{nam_microphone}, and earphone-type microphones\cite{clearbuds}.

\subsubsection{Unidirectional microphones}
Unidirectional microphones, such as those found in pin microphones and headsets, bear one of the most readily available noise reduction technologies today. 
Noise reduction is achieved by blocking the direction of vibration of the diaphragm, from the back side and restricting it to one side, thereby narrowing the directivity to 180 degrees; note that the diaphragm is found inside the microphone\cite{A-century-of-microphones}. This allows for strong recording of sound in a limited direction relative to the microphone, which is important when multiple speakers are speaking at the same time, such as in a panel discussion. However, since these microphones do not limit the distance of sound, background noise cannot be removed by the microphone by itself, making them unsuitable for use in noisy environments.

\subsubsection{Array microphones}
Array microphones are equipped with multiple microphones, and beamforming\cite{bss_review1} is used to narrow down the direction of arrival of sound by taking advantage of the time difference between the arrival of emitted sound at each microphone. 
This allows the selection of the speaker of an utterance. However, since multiple microphones which are arranged either in a circular fashion or in a straight line, are required to narrow down the direction of sound, using this system in a wearable device is challenging. Furthermore, even if the direction of arrival of sound can be estimated, it is not possible to determine whether it is background noise or not, requiring post-processing to deal with background noise.

\subsubsection{Throat microphones}
A throat microphone uses a piezoelectric element to convert the vibrations that appear on the surface of the neck when a speech is uttered \cite{Throat-microphone}. By attaching the microphone onto the neck and acquiring only the surface sound, only the wearer's voice is collected; environmental noise cannot cause the piezoelectric element to vibrate sufficiently, resulting in a high immunity to background noise. However, because the device must be worn tightly around the neck, noise is generated by movements such as head shaking or nodding. Moreover, because the voice travels through the skin, the formants necessary for speech recognition are deficient, and thus post-processing is required to ensure audible and accurate speech recognition\cite{throatmic1, throatmic2}.

\subsubsection{NAM microphones}
A NAM microphone works when it is in direct contact with the skin behind the ear similar to a pharyngeal microphone; it is an audio input device wherein an omnidirectional microphone is directly attached to the skin through a silicon \cite{A-century-of-microphones}. Similar to a pharyngeal microphone, a NAM microphone greatly reduces the effect of ambient sound\cite{nam_microphone}, but nodding and other sounds become noise. However, unlike pharyngeal microphones, NAM microphones acquire sound by being worn behind the ear, although both microphones require post-processing such as speech enhancement\cite{NAM0, NAM1, NAM2}.

\subsubsection{Earbuds with microphone}
An earpiece-type microphone that is inserted into earphones has been proposed, to achieve clear speech input, with earphones attached to both ears. Earbuds are also used to collect health information by monitoring exercise and biometric data \cite{earbuds_sensing}. Earpiece microphones are beamformer systems involving the earphones mounted on both ears, allowing selective acquisition of the wearer's voice, moreover, the hardware is open source, facilitating its faster development \cite{opensource_earbuds1}. Machine learning methods have also been proposed to reduce noise from speech detected in both ears\cite{clearbuds}. However, these methods are difficult to implement for voices that are not produced clearly, such as whispers.

\subsection{Software approaches for noise reduction: speech enhancement and blind source separation}
In the field of speech signal processing, the use of BSS \cite{bss_review1, bss_survey_sawada_ono_kameoka_kitamura_saruwatari_2019} and speech enhancement\cite{enhancement_review1, enhancement_review2} have long been proposed. In BSS, the resolution of the space where voices propagate is increased by increasing the number of microphones \cite{bss_microphone_work_well_with_numbers}, and the independence of multiple speakers from each other is used as a criterion for source separation. 

In BSS, various methods have long been proposed such as increasing the number of microphones to increase the resolution of the space where voices propagate \cite{bss_microphone_work_well_with_numbers}, separating sound sources based on independence \cite{IVA_Independent-vector-analysis}, and decomposing matrices into lower dimensional matrices\cite{NMF, fastnmf}. Other approaches \cite{irlma_kitamura2015efficient, idlma_makishima2019independent} combine these methods with deep learning. Furthermore, methods for speaker separation and speech enhancement that operate on small models have been proposed \cite{Wave-u-net, sepformer, denoiser} and they work in real-time. Real-time sound source separation involves a process known as short-time Fourier transform (STFT), which achieves a quick decomposition time. This process, however, reduces the frequency resolution necessary for high-quality synthesis. To address this issue, learning methods have been introduced. These methods propose an evaluation function designed to maximize sound quality, thereby enhancing synthesis quality\cite{koisumi_enhancement}.

However, these machine learning-based methods demonstrate a limited generalization performance because they are based on specific English speaker data sets or specific noise data sets. For example, if a speaker is placed in front of a person's mouth and the exact same voice is played, the voice from the speaker will be misinput. Also, because the system is optimized for normal speech data, there are deviations from data in real environments, such as input from whispered voices or from whispered voices produced in a noisy environment.

\subsection{Silent speech interface for speech communication}
Silent speech, which facilitates interaction with a non-vocal input, such as lip image recognition as well as with speech, has been proposed \cite{silentspeech_denby, An_Introduction_to_Silent_Speech_Interfaces}. Silent speech uses not only the voice emitted from the vocal cords, but also lip image \cite{lip-interact,liptype, liplearner}, ultrasound image\cite{DNN-Based-Ultrasound-to-Speech,ultrasound1,sottovoce}, myoelectricity\cite{EMG0_wand, EMG1_wand, AlterEgo_emg4}, capacitance\cite{TongueBoard, silentspeller}, acceleration\cite{derma_ahs, SilentMask, jawsense}, strain\cite{E-mask}, magnetism\cite{endophasia}, EEG\cite{EEG1, EEG2_BCI}, and other modalities to measure the human activities leading to speech recognition. Since these methods do not emit sound, they are suitable for use in noisy environments (e.g., where many people are talking) 

However, silent speech input has lexical challenges in that it is not conversational and it faces interaction challenges such as wearability and hands-free input. SilentSpeller \cite{silentspeller} demonstrates a high performance in terms of vocabulary but is not very expressive given that the current vocabulary for speech recognition is 9 million words. Meanwhile, a large lip image data is available for lip image input, that can be used in dark environments or as scalable command recognition \cite{liptype}, but the limitations of the commands remain unclear. Moreover, users must face their smartphones to speak, making hands-free input difficult, which is possible with voice. Myoelectricity is envisioned for use in actual conversations, such as synthesizing speech from silent speech \cite{emg_array}. However, myoelectricity requires the attachment of myoelectric array electrodes onto the surface of the face, making it unsuitable for prolonged use, and it does not take into account the effects of walking and other movements.

Silent speech with breathing has also been proposed in the speech modality \cite{fukumoto}, but it requires learning the interaction of breathing in; moreover, its vocabulary is smaller than that of the touch sensors and updated images described above. Input by non-audible murmur has also been proposed\cite{NAM0, NAM1, NAM2}, but it has the problem that noise is generated by the user's natural movements, such as nodding or turning around, as well as the noise produced by touching the NAM microphone, are the identified problems.

\begin{figure*}[htbp]
 \begin{center}
  \includegraphics[width=\textwidth]{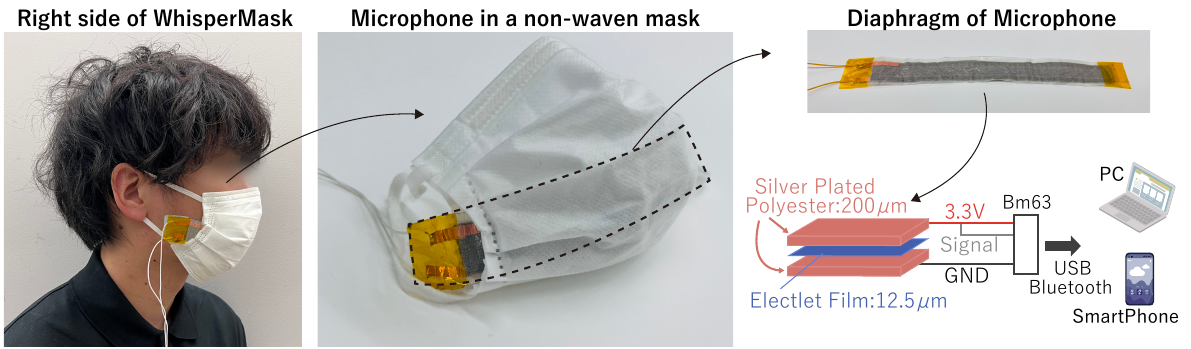}
  \caption{Overview of WhisperMask as a masked microphone. WhisperMask is a mask-type microphone that allows for hands-free, non-obtrusive input (right). The microphone is sandwiched between the fabric of two non-woven masks (center). The diaphragm is connected to a microcontroller and can be used on a PC or smartphone via USB or Bluetooth (right).}
  \label{fig:whispermask_device}
 \end{center}
\end{figure*}

\subsection{Mask-type interface}
The increased use of masks during the COVID-19 pandemic has led to a surge of interest in exploring their potential as wearable interfaces. While masks conceal facial expressions thereby posing a challenge, researchers have proposed methods to capture and present expressions using photo reflectors, capacitive touch sensors, LEDs, and displays \cite{unmasked, mascreen, TransEmotion, TexonMask}. Furthermore, masks have been adapted as wearable interfaces for various interactions involving the face and mouth, including breath detection \cite{ibuki, ReliableBreathing, Respiration}, eye tracking \cite{SleepMask}, mouth shape recognition \cite{yadori, mouthgesture}, and mask attachment/removal detection using the straps \cite{masktrap}.

Notably, masks have been reported to cause voice attenuation, making speech perception difficult \cite{mask-is-difficult-to-speech-recognition1, mask-is-difficult-to-speech-recognition2}. To address this issue, researchers have proposed embedding sensors in masks to enable silent speech recognition \cite{SilentMask, E-mask}. However, these approaches primarily focus on recognizing specific commands for communication with smart assistants such as Alexa, limiting their applicability for broader speech recognition and conversation capabilities.

While products such as HushMe \cite{Hushme} and Mutalk \cite{mutalk} offer voice isolation in noisy environments by fully sealing around the mouth, they are bulky and are intended primarily for gaming applications, limiting face-to-face communication. Similarly, microphone devices with built-in ventilation fans \cite{LGPuriCare2024}, although lightweight and sleek, necessitate constant fan operation to ventilate the sealed area, leading to considerable noise pollution.

Our proposed solution, WhisperMask, addresses the need for voice isolation with the use of a single lightweight mask, thus offering a practical alternative. Importantly, the hands-free and non-obtrusive nature of WhisperMask makes it particularly well-suited for environments where mask usage remains prevalent, such as operating rooms\cite{ASR_in_operating_room}, clean rooms, or other noisy settings where clear communication is crucial. By effectively suppressing background noise while preserving user comfort and privacy, WhisperMask enables reliable voice interactions in these challenging scenarios.

\section{WhisperMask, an ELECRET CONDENSER MICROPHONE(ECM)-based mask-type microphone}
./


 We propose a wearable mask-type microphone called WhisperMask, which allows for  input even in noisy environments. WhisperMask is designed based on the principle of the electret condenser microphone (ECM) and incorporates a vibrating diaphragm designed using conductive fabric and film.

\subsection{Principle: Electret Condenser Microphone}
WhisperMask is a microphone designed based on the ECM technology. An ECM consists of a power supply section and a vibrating diaphragm, where the diaphragm forms a capacitor with the electrode, giving rise to a microphone. When a sound causes the diaphragm to vibrate, the voltage of the capacitor changes, resulting in a weak voltage that is converted into sound through a field effect transistor(FET) and through analog-to-digital conversion. In this study, we propose a novel design of a vibrating diaphragm for ECM, enabling selectivity toward the speaker's voice.

\subsection{Design of WhisperMask}
The diaphragm of our microphone consists of a conductive fabric electrode and a dielectric plastic film, forming an electret condenser. The plastic film used is PFA (perfluoroalkoxy polymer), with a thickness of 12.5 $\mu$ m. For wind protection, an adhesive tape was applied around the film, and a 200 $\mu$ m silver-plated polyester cloth was fixed onto both sides of the tape. This configuration forms the structure of an electret condenser wherein the electret was sandwiched between two electrodes.

The diaphragm was connected to an FET, which outputs the audio signal. Compared with commercially available condenser microphones, the proposed microphone has a lower voltage value, which is amplified by a factor of 3 using FETs. Furthermore, the input signal is connected via Bluetooth to a Bm63 audio system-on-a-chip (SoC), which is housed in a 6 mm x 5 mm housing and weighs less than 10g.

After producing the diaphragm and the measuring circuit, we embedded them in a breathable mask created from a mesh-type fabric. The mask was designed to bear slits to allow for the positioning and fixation of the diaphragm.

\subsection{Patterns of the diaphragm}
WhisperMask is a mask-type microphone, which is used by wearing a mask. However, if the sensor becomes heavy, the mask will shift and fall off, resulting in decreased recognition. Therefore, in this study, we prepared three sizes of microphones (20 mm×4 mm, 20 mm×2 mm, and 10 mm×2 mm , Fig. \ref{fig:whispermask_diaphragm_pattern}), and we embedded them in a mask for assessment (Fig. \ref{fig:whispermask_device}).



\begin{figure*}[htbp]
    \centering
    \includegraphics[width=\textwidth]{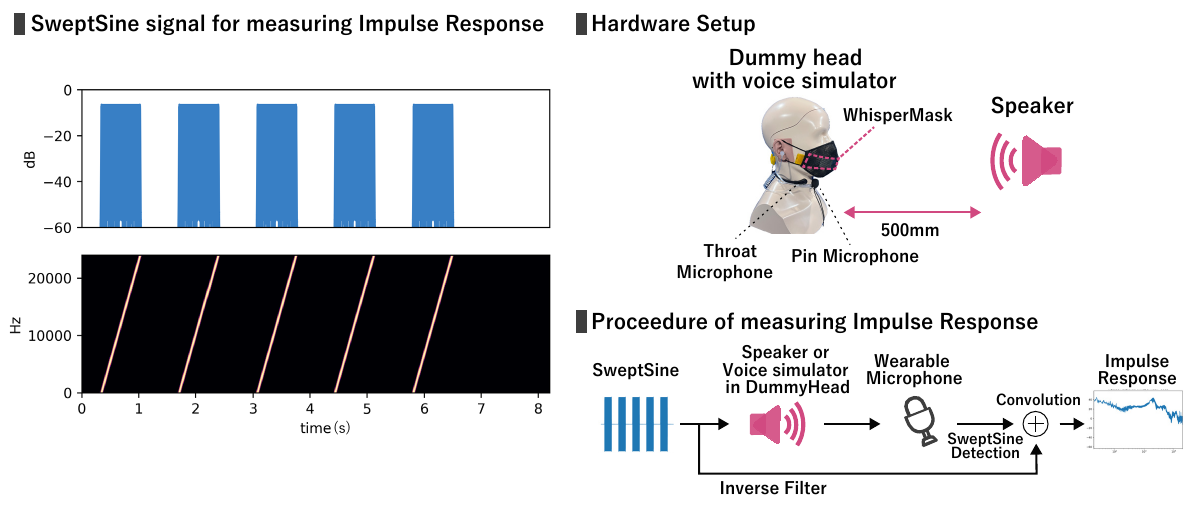}
    \caption{Swept-sine wave for measuring impulse response. Five Swept-sine of length 65536 are generated(left). A dummy head with a voice simulator is provided and the microphones are worn for measurement. The speaker is placed 500 mm away from the dummy head(upper right). The impulse response is calculated by convolving the signal obtained by preprocessing to detect Swept-sine with the inverse filter of Swept-sine. (lower right)}
    \label{fig:whispermask_proceedure_of_measuring_inpulse_response}
\end{figure*}

\section{Frequency Characteristic of WhisperMask}
\label{sec:eval_sweptsine}
Frequency characteristic is one of the most critical aspects of a microphone, as it determines the microphone's ability to capture specific frequencies and to accentuate or attenuate certain frequency bands. For example, in voice recognition, which is crucial in voice-based interactions, human speech predominantly falls within the range of 5 kHz, and a microphone that does not capture this frequency range would make voice recognition extremely challenging. Additionally, it is important to consider how clearly a signal stands out against noise, which is measured in terms of SNR. 
A low SNR value means that a significant part of the obtained signal is noise, making it difficult to capture the desired signal, such as the speaker's voice.

\subsection{Impulse response with swept-sine}
\subsubsection{Characteristic of the swept-sine signal}
Frequency characteristics can be measured using the microphone's impulse response, which in turn can be measured using various methods, such as using an impulse wave or white noise. However, the widely employed method involves swept-sine, a continuous sine wave that sweeps from low to high frequencies\cite{swept-sine}. Swept-sine measurement provides information across a wide frequency range using a single signal. It is robust against noise and relatively easy to measure. Furthermore, the deterministic properties of a swept sine signal make it less noisy compared with measurement methods involving white noise. By taking measurements repeatedly, random noise can be reduced, enabling the acquisition of accurate and reproducible values. The waveform obtained from swept-sine represents the convolution of the system's impulse response and the swept-sine itself. To obtain the microphone's impulse response from the acquired signal, an inverse convolution filter can be applied during the generation of a swept-sine signal (Fig. \ref{fig:whispermask_proceedure_of_measuring_inpulse_response}, right).

\subsubsection{Design of the swept-sine wave for measurement}
A swept-sine is capable of resolving different frequencies based on the length of a sample, with longer samples providing higher a frequency resolution. In this study, we used a sample length of 65536 points to achieve a frequency resolution of less than 1 Hz when the sampling frequency was set at 44.1 kHz. The frequency range of the swept-sine was limited to up to 22.1 kHz, which corresponds to the upper limit of human auditory perception. As mentioned earlier, repeating the swept-sine measurement improves the SNR; thus, conducting multiple measurements is crucial.
Therefore, a swept-sine consisting of 65535 samples was created and the process was repeated five times to generate a single wav file (Fig. \ref{fig:whispermask_proceedure_of_measuring_inpulse_response} left). 
However, to identify the starting positions of each swept-sine, a buffer of the same length as the swept-sine was inserted between consecutive swept-sines. By including these gaps, the input consisted of five repetitions of the swept-sine, which was then repeated 10 times, resulting in a total of 50 measurements of the impulse response. This averaging process was carried out to assess the system's noise robustness. Finally, we use 1/3 octaveband averaging \cite{octaveband-smoothing} to smoothen the impulse response.

\begin{figure*}[htbp]
    \begin{tabular}{cc}
      \begin{minipage}[t]{0.45\hsize}
        \centering
        \includegraphics[scale=0.05]{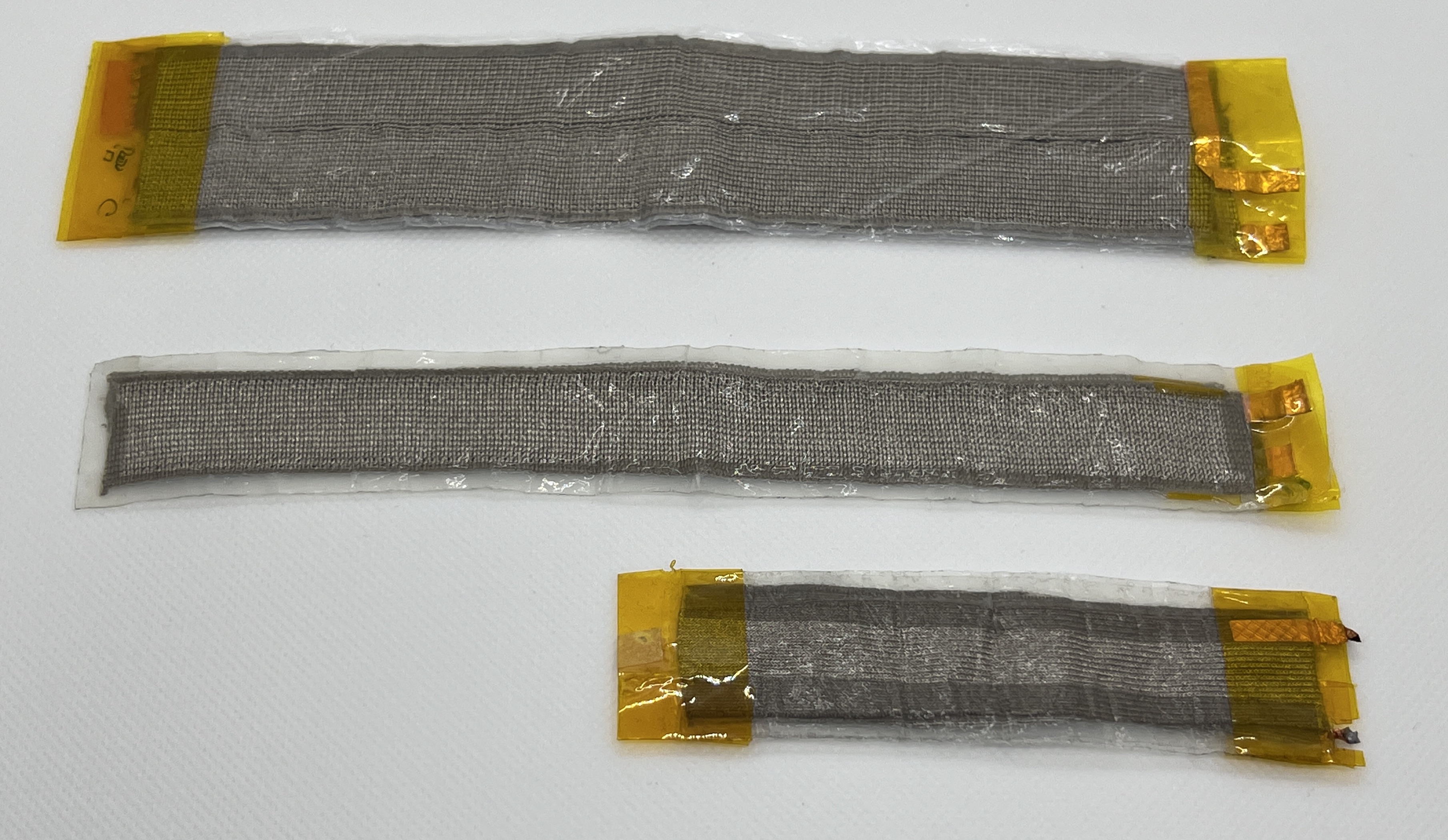}
        \caption{Sizes of the WhisperMask diaphragm: \\ 20 mm×4 mm(upper), 20 mm×2 mm(center), and 10 mm×2 mm(lower)}
        \label{fig:whispermask_diaphragm_pattern}
      \end{minipage} &
      \begin{minipage}[t]{0.45\hsize}
        \centering
        \includegraphics[keepaspectratio, scale=0.5]{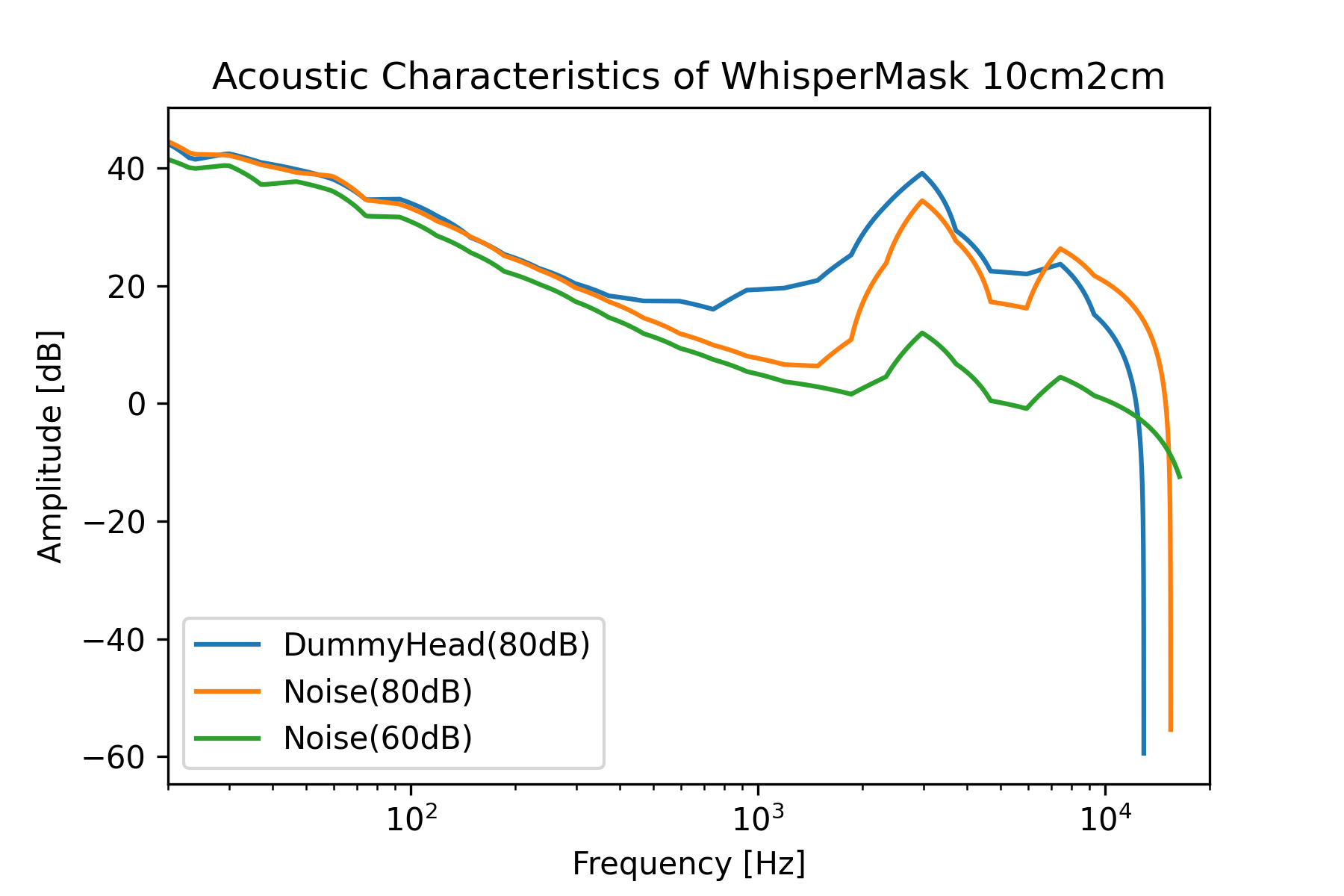}
        \caption{Impulse response of WhisperMask(10 mm×2 mm)}
        \label{whispermask_IR_10cm2cm}
      \end{minipage} \\
   
      \begin{minipage}[t]{0.45\hsize}
        \centering
        \includegraphics[keepaspectratio, scale=0.5]{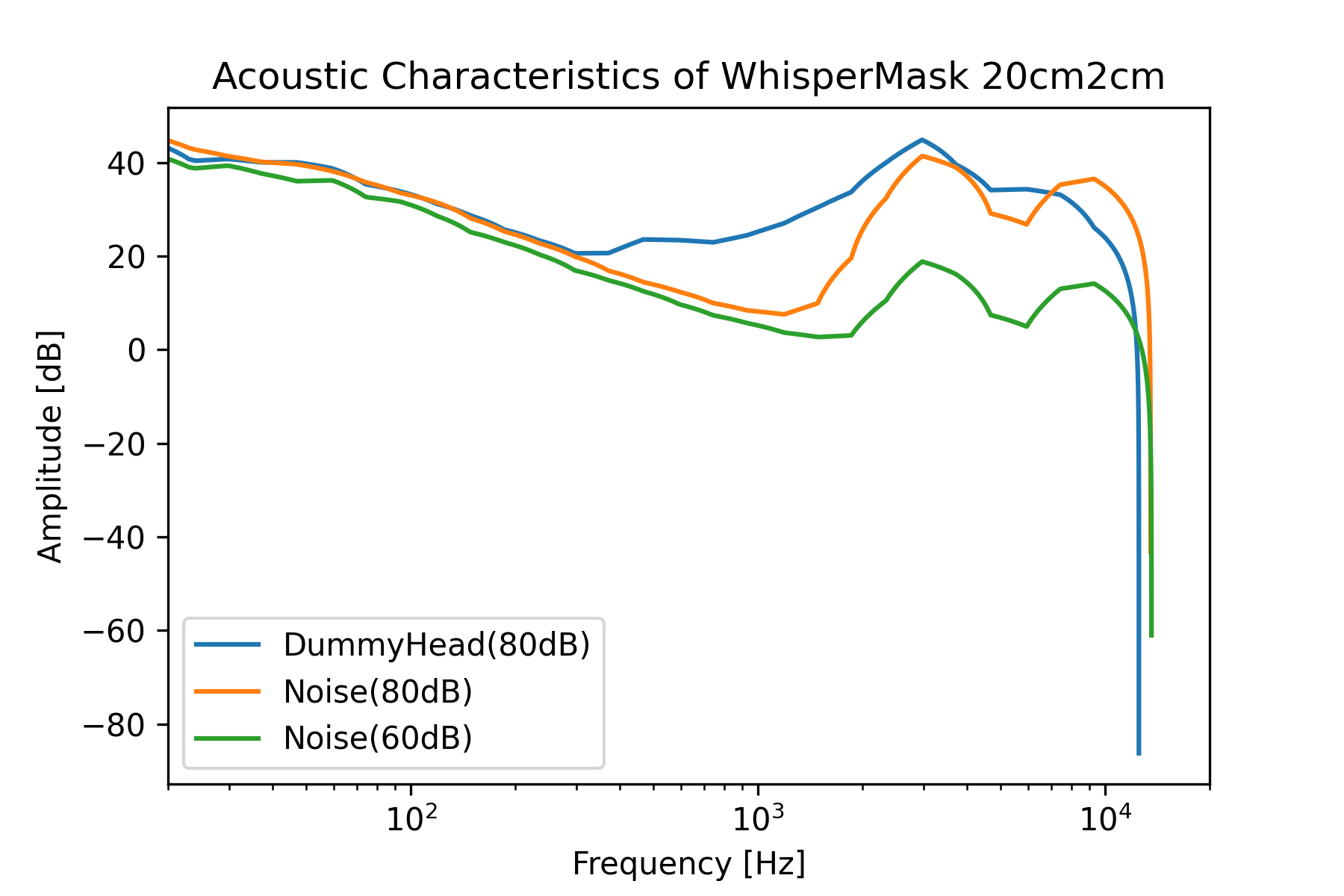}
        \caption{Impulse response of WhisperMask(20 mm×2 mm)}
        \label{whispermask_IR_20cm2cm}
      \end{minipage} &
      \begin{minipage}[t]{0.45\hsize}
        \centering
        \includegraphics[keepaspectratio, scale=0.5]{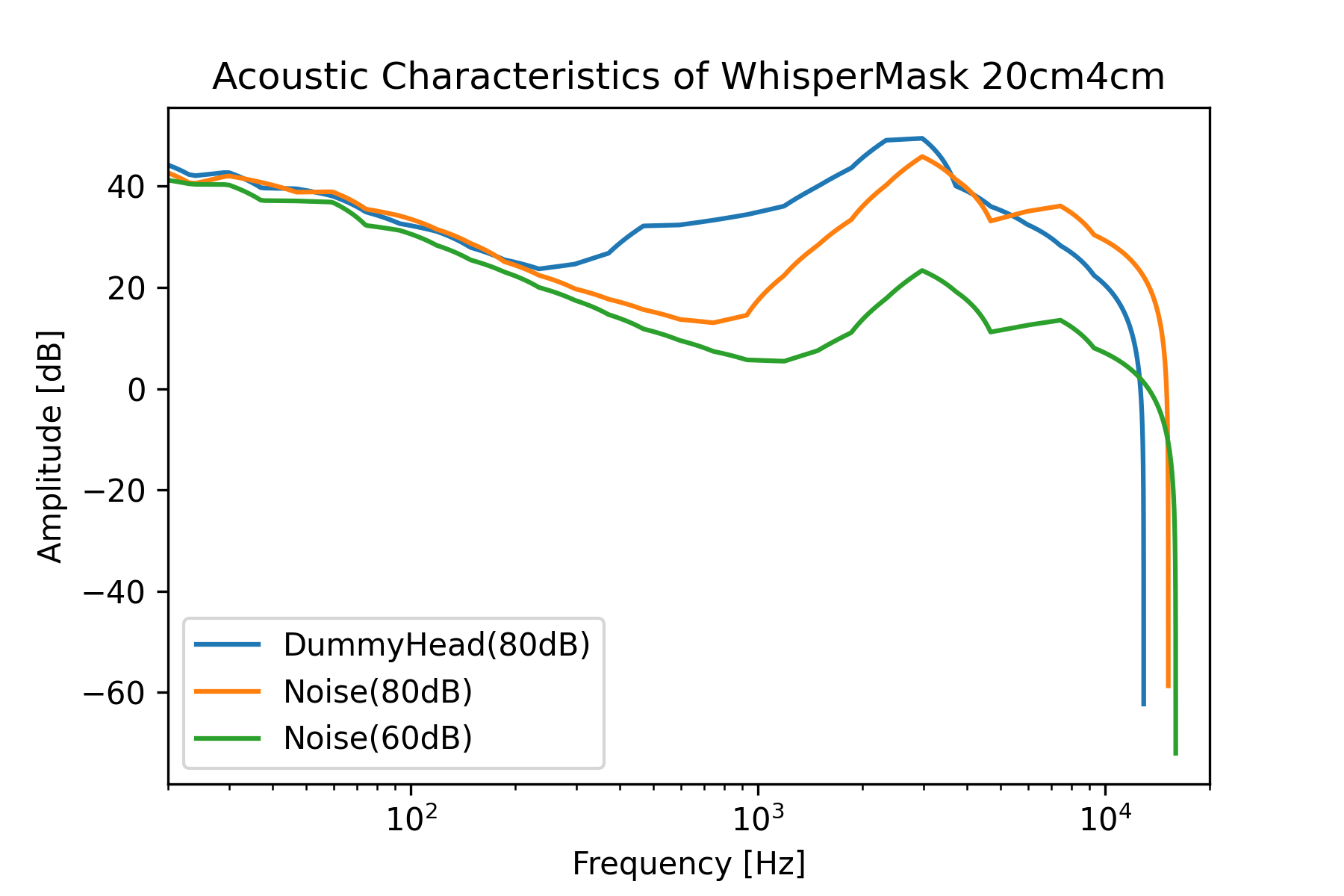}
        \caption{Impulse response of WhisperMask(20 mm×4 mm)}
        \label{whispermask_IR_20cm4cm}
      \end{minipage} 
    \end{tabular}
\end{figure*}

\subsection{Experimental condition}
\subsubsection{Environmental setup for measuring impulse response}
For the impulse response measurement, a dummy head (SAMAR4700M) equipped with a mouth simulator was used to mimic human speech output. The SAMAR4700 complies with the international standards IEC 60318-7 and ITU-T Rec.P51 for head shape and mouth simulator, enabling the simulation of human speech for measurements. The dummy head was positioned on a tripod 40 mm above the floor, and a speaker that replicates the emitted sound was fixed 50 mm in front of the dummy head.

The acoustic measurements were conducted in an electromagnetic anechoic chamber to minimize interference from electromagnetic waves and certain sound waves. The room was treated with porous materials to absorb sound, resulting in a room noise of 28.8 dB.

\subsubsection{Input signal}
During the measurements, a dummy head simulating human speech and a speaker imitating external noise were set up. To avoid distortion of the output sound, 80 dB swept-sine was output as a human voice from a mouth-simulating speaker. Two swept-sines (80 and 60 dB) were output as background noise from the external noise speaker (Fig. \ref{fig:whispermask_proceedure_of_measuring_inpulse_response} right upper). These sound levels were measured with a precision noise system, and the obtained values fell within an error of 0.1 dB.

\subsection{Swept-sine detection}

The output of the swept-sine is susceptible to variations in frequency characteristics depending on the properties of the receiving microphone. In other words, it is impossible to acquire the full range of frequency bands, and in some cases, only a partial representation is obtained. However, to accurately measure the impulse response, determining the timing at which the swept-sine signal begins is crucial. To estimate the start and end times of the swept-sine from the information obtained within certain frequency bands, the following approach was employed. Firstly, the swept-sine was decomposed into individual frequencies at 1 kHz intervals, and the envelope was acquired. As the swept-sine was output five times with an interval of one pulse, five rising edges appeared in the frequency regions where the swept-sine was well represented. By capturing these, five points were recorded. Dividing the frequencies at intervals of 1 kHz, a maximum of 22 points (22 kHz/1000) was recorded. Using these points, a linear regression was performed, with the point at frequency 0 representing the start time and the point at frequency 22000 representing the end time.

\subsection{Result for impulse response}

The impulse response results are shown in Fig. \ref{whispermask_IR_10cm2cm}, \ref{whispermask_IR_20cm2cm}, \ref{whispermask_IR_20cm4cm}. The blue line represents the impulse response of the assumed human speech, played from the mouth-simulating speaker of the dummy head, with 80 dB swept-sine. The orange and green lines are both assumed to be noise and were output at 80 dB and 60 dB, respectively, from the external speaker.

In each of the three patterns, especially in the frequency band between 200 Hz and 5 kHz, the output of the dummy head was approximately 10 dB higher than the outside noise, indicating that when the same waveform at the same sound pressure is input inside (dummy head) and outside (noise simulating speaker) the microphone, the microphone captures the inside sound in more easily, that is, it reduces noise.

   

 \begin{figure*}[htbp]
    \centering
    \includegraphics[scale=0.7, trim=0cm 0.5cm 0cm 0cm]{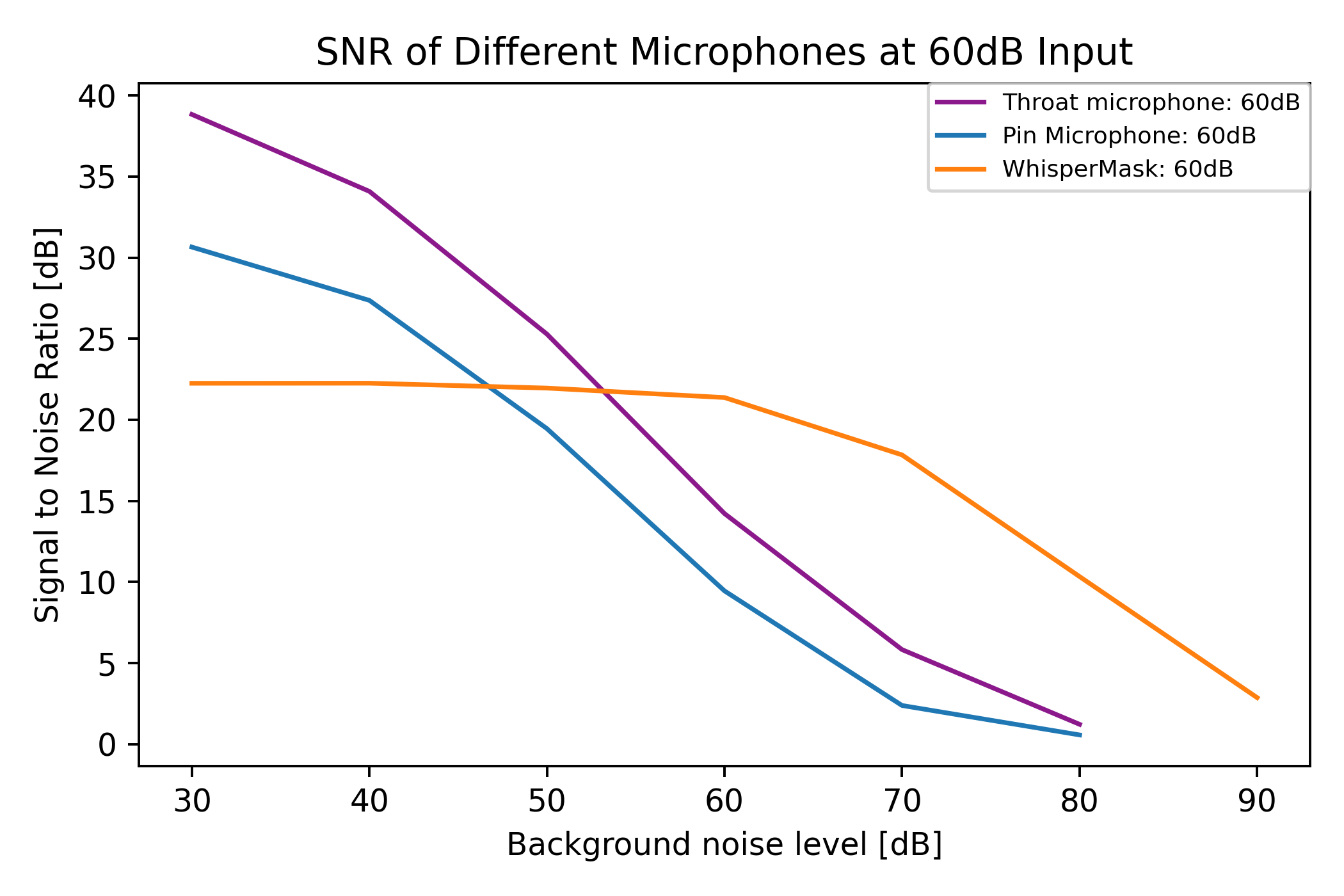}
    \caption{SNR result for different microphones at 60 dB input}
    \label{fig:result_snr_60 dB}
\end{figure*}
\section{Evaluation: Noise suppressing Microphone }
\label{sec:charactristic_noise_suppressive_microphone}
\subsection{Evaluating the effect of noise suppression: SNR measurement}
The proposed microphone can capture a speaker's voice in noisy environments. To demonstrate this, the SNR was measured. The SNR was computed by recording the output (N) of the microphone without supplying input signals from the speaker and by recording the output (S) when input signals were generated. The root mean square (RMS) values were then calculated for each microphone output, and the SNR was determined using \begin{math}
20log10(S_{RMS}/N_{RMS})    
\end{math} \cite{snr_novel}. SNR evaluation is commonly used to assess the performance of microphones in noisy environments and is also utilized in array microphones \cite{snr_reverbing}.

\subsection{Environmental setup for SNR measuring}
For SNR measurement, a dummy head(SAMAR4700M) equipped with a mouth simulator was used to mimic human speech output. The SAMAR4700 complies with the international standards IEC 60318-7 and ITU-T Rec.P51 for head shape and mouth simulator, enabling the simulation of human speech for measurements. The dummy head was positioned on a tripod 40cm above the floor, and a speaker that replicates the emitted sound was fixed 50cm in front of the dummy head.

Acoustic measurements were conducted in an electromagnetic anechoic chamber to minimize interference from electromagnetic waves and certain sound waves. The room was treated with porous materials to absorb sound, resulting in a room noise of 28.8 dB.

The input signal used for the measurement was a 20 Hz--20 kHz swept-sine to also perform calculations in the full range. The sound pressure of the signal output from the dummy head was 60 dB, which is close to that of human speech. Environmental noise was output from an external speaker in 10 dB increments from 30 dB to 90 dB, with an upper limit of 90 dB based on the characteristics of the speaker's output.

Three devices were used for measurements: WhisperMask (10 mm x 4 mm), pin microphone(PinMic), and throat microphone(ThroatMic). The earbuds with microphone (Airpods) were optimized for human voices and could not input noise-like waveforms similar to those of the swept-sine. Each microphone was mounted on a dummy head in an ideal position on the dummy head.

\subsubsection{SNR results for noisy environment}
 The inputs in the noisy environment were as follows: the SNR for PinMic and ThroatMic decreased as the ambient noise increased, whereas that for WhisperMask hardly changed from 30 dB (SNR: 22.2) to 60 dB (SNR: 21.3) of ambient noise. SNR was calculated as the ratio of the power of the signal to the power of the noise:
\begin{math}
SNR=20log10(S_{RMS}/N_{RMS})    
\end{math}.
 And since the input signal(\begin{math}S_{RMS}\end{math}) is almost constant at 60 dB swept-sine, from 30 dB to 60 dB. WhisperMask picked up almost no ambient noise because \begin{math}N_{RMS}\end{math} is nearly constant.

The impact of noise on WhisperMask's performance was greater when the external noise was higher than 70 dB; at 70 dB, the SNR for WhisperMask was 17.83, which was 10 dB higher than the SNR values for PinMic(2.3) and for ThroatMic(5.83).

\section{Evaluation: Quality of THE Recorded Voices} 
Widely used for voice input and calls, microphones require not only noise resilience but also high-quality sound during recordings. We evaluated the sound quality in noisy conditions to compare WhisperMask not just against existing microphones but also against the performance of conventional microphones following the use of a noise reduction software.

\begin{figure*}[htbp]
  \begin{minipage}[b]{0.45\linewidth}
    
    \includegraphics[scale=0.25]{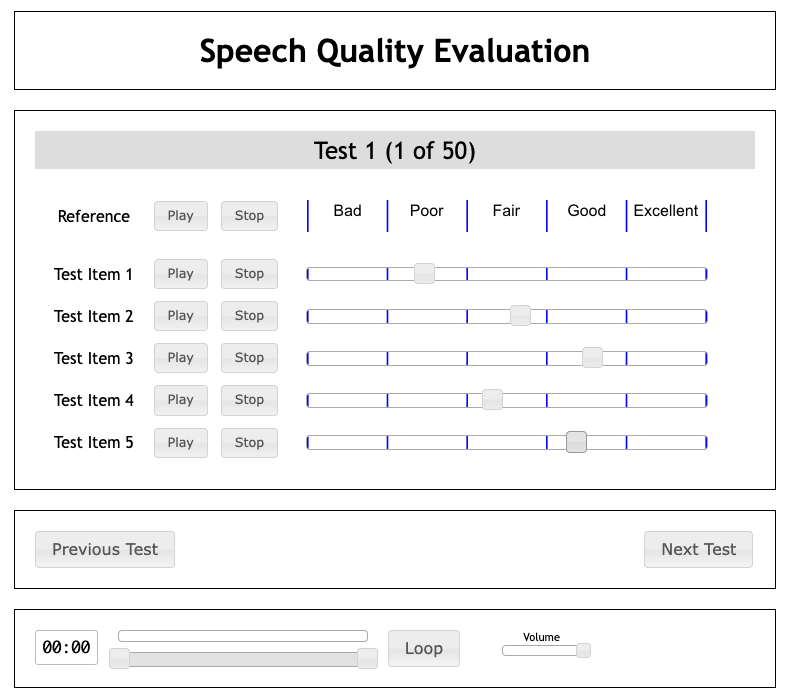}
    \caption{The WebUI used to evaluate the quality of the recorded audio clips; the metrics were based on MUSHRA\cite{mushra-ITU-R}. The participants rated each audio clip on a scale of 0 to 100. The reference audio clip (top row) had a high quality and was used as a criterion for selecting responses. Four of the five test items are recordings captured in a noisy environment by each device. One test item is the same as the reference, any participants who rated this test item as having lower quality than the reference audio will be judged as less faithful respondents.}
    \label{mushra_webui}
  \end{minipage}
  \hspace{0.04\columnwidth} 
  \begin{minipage}[b]{0.45\linewidth}
    
    \includegraphics[scale=0.5]{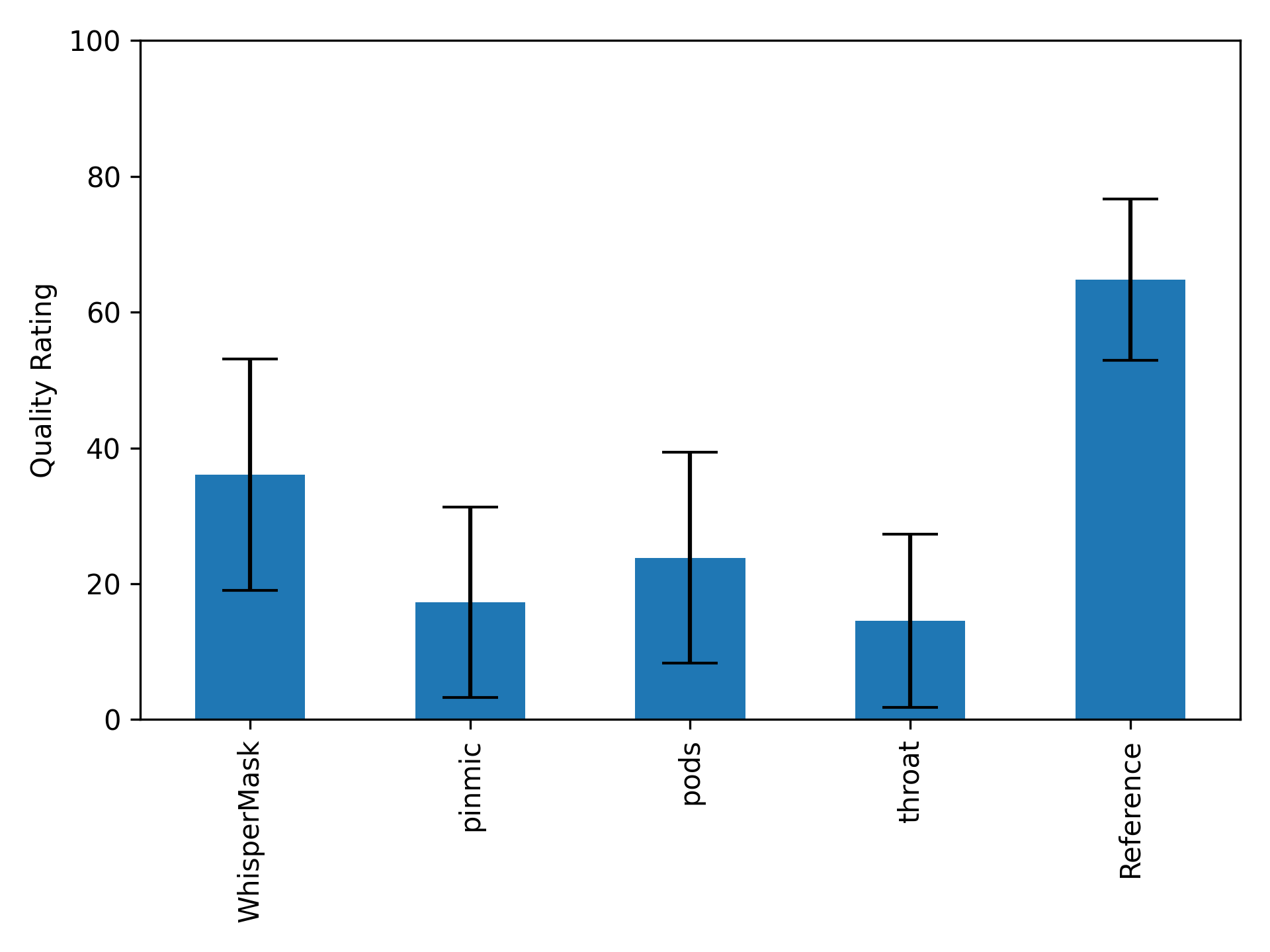}
    \caption{Assessment results for the quality of audio recorded by WhisperMask and other devices in an environment with 80 dB background noise. “Pinmic,” “pods,” and “throat” in the x-axis represent unidirectional pin microphones, Airpods Pro2, and throat microphones, respectively. The reference is a clear sound recorded in a quiet environment(30 dB) with a pin microphone.}
        \label{result_mushra_original}
  \end{minipage}
\end{figure*}

\label{sec:mushra_in_Noisy_Environments}
\subsection{Record conditions}
The audio recordings used in this study were of five phrases extracted from the Text Entry dataset\cite{MacKenzie_Soukoreff}, and read aloud by four proficient English speakers (one male and three females). These speakers consented 
consented to participate in this study and become subjects for data collection, for which they received \$20. The noise level during the recording was 80 dB, and the four microphones used for recording were WhisperMask, unidirectional pin microphone, earbuds with microphone (Airpods Pro2), and throat microphone. The reference audio is a clear sound recorded in a quiet environment (30 dB) with a pin microphone.


To compare with software-based noise reduction, we created recordings by applying noise removal software to the captured audio clips. There are two main approaches to noise removal in noisy environments: noise removal and speech enhancement. These approaches are not always explicitly compared in speech research.
Therefore, for comparison, we choose one method from each approach that utilizes a learning-based embedding model for comparison. For speech enhancement, we used a denoiser \cite{denoiser}. Denoiser is an extension of U-Net \cite{u-net} and is trained on the noisy speech dataset \cite{denoiser_dataset-1, denoiser_dataset-2}. For speech separation, we use a waveformer \cite{waveformer}. A waveformer is a model that extends CNN to handle sequential data and is trained to synthesize and separate sounds from a selected speech dataset \cite{waveformer-dataset-1} and an environmental sound dataset \cite{waveformer-dataset-2}.

\subsection{MUSHRA: a metrics for evaluating audio quality}
For subjective evaluation of sound quality, the Multiple Stimuli with Hidden Reference and Anchor (MUSHRA) has been proposed. Compared with Mean Opinion Score(MOS), MUSHRA is defined in ITU-R BS.1534 and is evaluated on a scale of 0 to 100, allowing for the evaluation of subtle differences. Multiple test items were provided to the evaluators among these items is an "anchor" wherein the sound quality was intentionally reduced to provide a reference for the quality evaluation. This approach ensures consistent evaluations \cite{mushra-ITU-R}.

MUSHRA evaluations have been used in a wide range of applications in speech processing, including noise reduction \cite{mushra-ex1-denoise}, text-to-speech synthesis \cite{mushra-ex2-tts}, and voice transformation \cite{mushra-ex3-wesper}. It was also been used for evaluations by the participants of an online experiment \cite{webmushra}.

\begin{figure*}[htbp]
  \begin{minipage}[b]{0.45\linewidth}
        \includegraphics[scale=0.45]{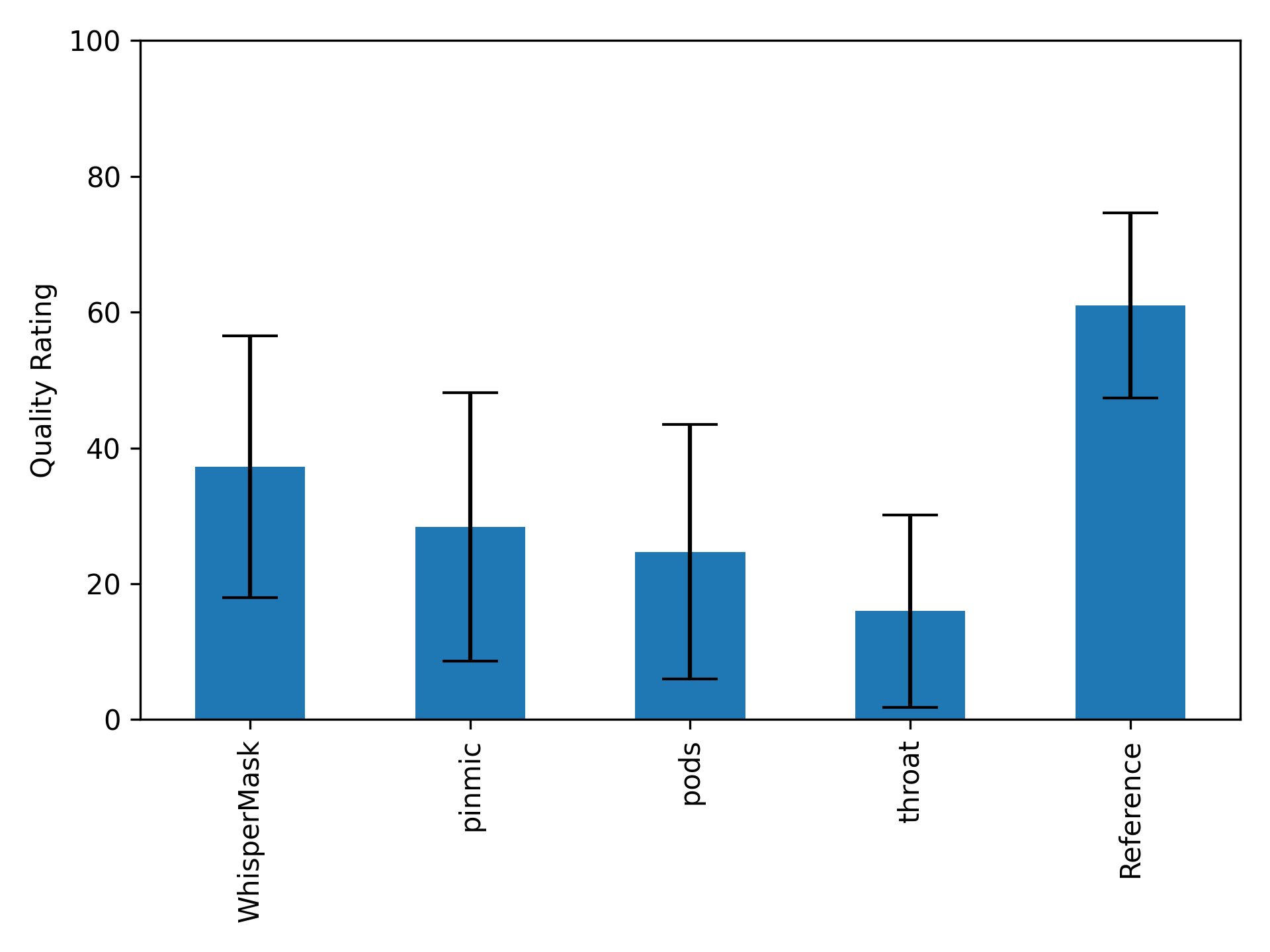}
    \caption{Audio quality of the recording captured by WhisperMask following post-processing with denoiser \cite{denoiser}, a speech enhancement software.}
    \label{result_mushra_denoiser}    

    \label{mushra_webui}
  \end{minipage}
  \hspace{0.06\columnwidth} 
  \begin{minipage}[b]{0.45\linewidth}
    
    \includegraphics[scale=0.45]{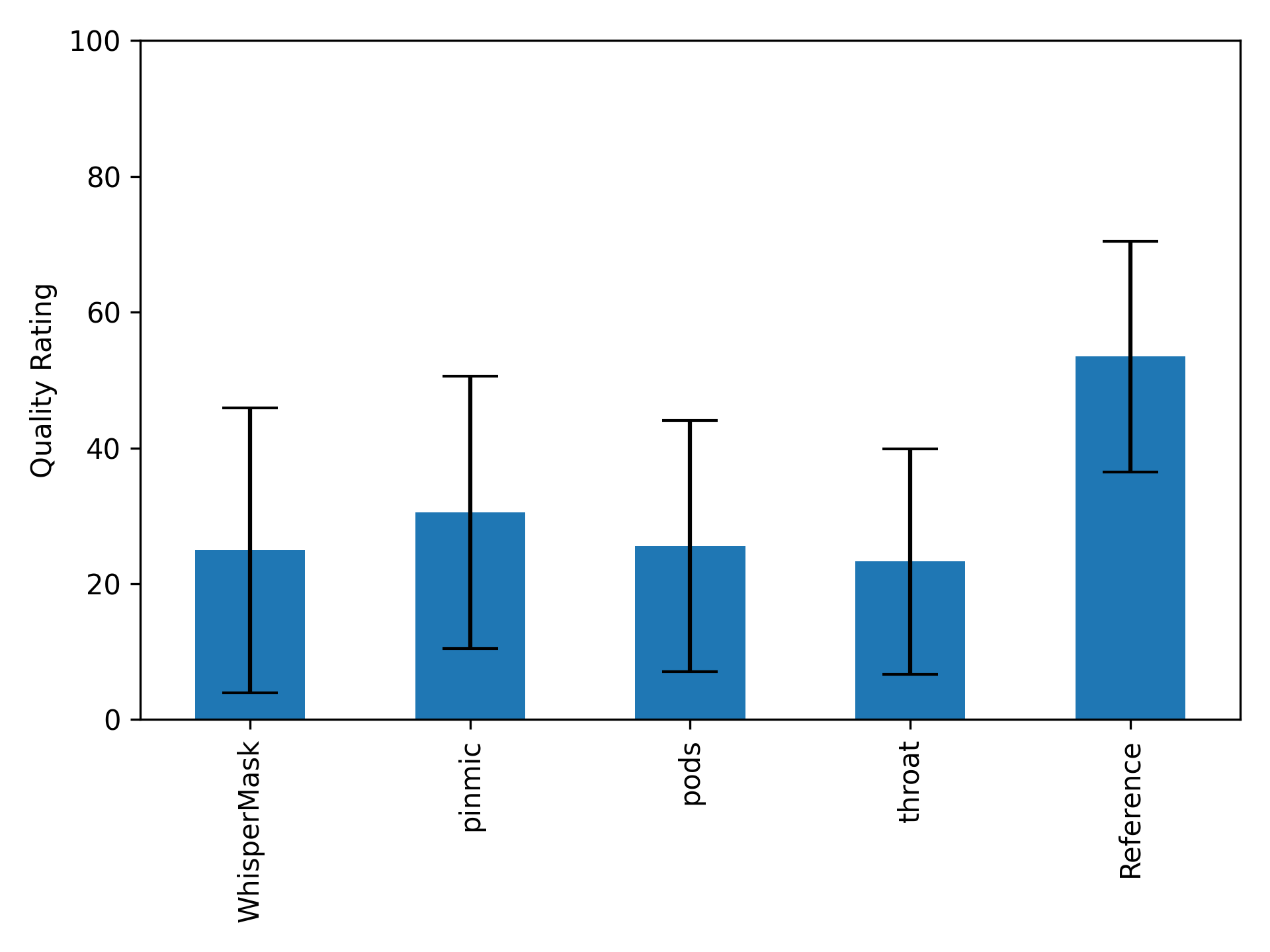}
    \caption{Audio quality of the recording captured by WhisperMask following post-processing with sepformer \cite{sepformer}, a speech separation software.}
        \label{result_mushra_sepformer}
  \end{minipage}
\end{figure*}

\subsection{Experimental proceedure}
We recruited 25 native English speakers (13 males and 12 females) aged 18 and above through Mechanical Turk. The audio quality was evaluated using WebUI, as shown in Fig. \ref{mushra_webui}.

In the WebUI, the participants would rate each audio clip on a scale of 0 to 100; a high-quality audio clip was provided as a reference. 
One of the five choices was the same as the reference, enabling the identification of less faithful respondents. The participants may listen to the sound clips as many times as they wished.
The average completion of the participants in the experiment was 49 minutes. They were compensated \$7 for their participation.

\subsection{Result of the audio quality evaluation}
The results are shown in Fig. \ref{result_mushra_original}. “Pinmic,” “pods,” “throat” represent unidirectional pin microphone, Airpods Pro2, and throat microphone, respectively. The t-test results show that WhisperMask was superior to Pinmic, pods, and throat at p=5.06E-22, 2.95E-10, and 1.50E-28 (p<0.05), respectively, and with corresponding effect sizes of 1.20, 0.75, 1.43 \cite{cohen1962statistical}.

\subsection{Comparison of sound quality with and without noise reduction}
For the experiment, we used MUSURA\cite{mushra-ITU-R} as a metrics, and we recruited 25 native English speakers (13 males and 12 females) aged 18 and above through Mechanical Turk.
The results are as shown in Fig. \ref{result_mushra_original}. As mentioned above, "pinmic", "pods", and "throat" represent unidirectional pin microphones, Airpods Pro2, and throat microphones, respectively. The t-test result in Fig. \ref{result_mushra_denoiser} show that WhisperMask was superior to pinmic, pods, and throat, at p=1.31E-4, 4.51-08, and 1.83E-22 (p<0.05), respectively with corresponding effect sizes of 0.455, 0.659, and  1.257 \cite{cohen1962statistical}. 

By contrast, when sepformer was used (Fig. \ref{result_mushra_sepformer}), pin microphone was superior to WhisperMask at p=0.02 in t-test (p<0.05) and 0.27 in effect size. The performance of the other devices (AirPods and throat microphone) did not statistically differ from that of WhisperMask (p=0.78, 0.45 in t-test).


Furthermore, the quality of recording captured by WhisperMask with and without denoiser or sepformer did not significantly differ (p=0.575>0.05), and it was better than the sound quality obtained using sepformer (p=8.74E-07<0.05; effect size 0.583

\section{EVALUATION: Speech Recognition Accuracy} 

Speech input is widely used not only in telephony but also for interactive tasks, such as operating smart assistants and interactive searching using speech recognition. In this study, we performed speech recognition using two noise-robust speech recognition methods. Whisper \cite{openai-whisper} employs an encoder–decoder transformer model trained through supervised learning.


\begin{figure*}[htbp]
    \centering
    \includegraphics[width=\textwidth, trim=0cm 0.3cm 0cm 0.5cm]{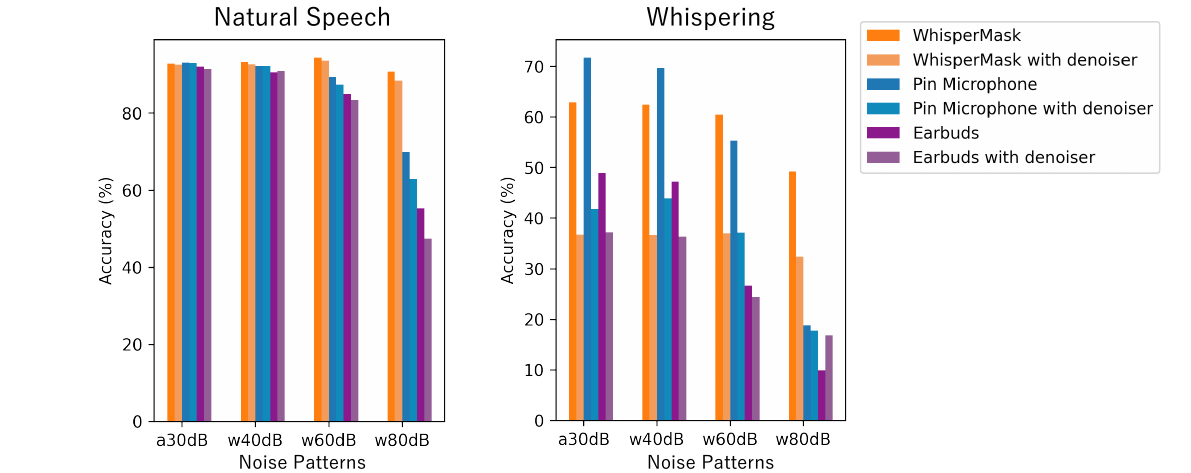}
    \caption{Results for speech recognition in environments with different noise levels. a30dB indicates a recording environment without white noise, and w40dB, w60dB, and w80dB indicate a recording environment with a white noise output of 40, 60, and 80 dB, respectively. In natural speech (left), recognition accuracy improved in the following decreasing order: WhisperMask followed by pin
microphone and then by earbuds. A difference of approximately 30\% was observed for whispered speech in an environment with 80 dB background noise (right).
Noise reduction by Denoiser \cite{denoiser} did not result in considerable changes in the recognition accuracy for natural speech but it significantly reduced that for whispered speech.
}
    \label{fig:software_vs_hardware_wer_result}
\end{figure*}

\subsection{Data Collection}
Data were collected from 9 participants (mean age 26.2 years; 4 males and 5 females). The participants were asked beforehand to rate their English proficiency on a 5-point scale, with 3 being the mean. During data collection, each participant read 20 pre-prepared phrases collected from the Mackenzie and Soukoref dataset \textbf{\cite{MacKenzie_Soukoreff}.}

In the experimental environment, white noise was varied at 40, 60, and 80 dB (denoted as w40, w60, and w80 dB in Fig. \ref{fig:software_vs_hardware_wer_result}, respectively). a30 dB indicates the noise level in the space when no white noise was being played. During the experiment, two methods of delivery were used: natural speech and whispered speech. The three microphones evaluated were WhisperMask, pin microphone, and earbuds with microphone (AirPods Pro2).

The background noise was played from a laptop and was output in a stereo; it was measured using a noise system to ensure that the desired sound pressure was reaching the user’s mouth and then adjusted to a difference of 0.5 dB or less. 
Measurements were taken in a soundproof room where the normal noise level was 30 dB. In each microphone and noise environment, two types of speech were used: natural speech and whispered speech

\subsection{Analysis}
In speech research, speech enhancement methods have been proposed to reduce background noise \cite{enhancement_review1,enhancement_review2}. In recent years, real-time noise reduction systems have become highly accurate \cite{enhancement_review1,enhancement_review2}, and it is already possible to apply noise reduction in microphones, such as pin microphones and AirPods, to obtain the desired audio quality. This study examines how much speech recognition accuracy can be improved relative to the audio quality obtained after applying real-time denoising to speech recorded with a pin microphone or AirPods. In a noise-free environment (30 dB), the participants’ average recognition rate for natural speech recorded using pin microphones and AirPods was over 90\%. The voice recognition rate for throat microphone was significantly lower at 64\% compared with that for the other devices. This discrepancy was attributed to improper fitting for some participants, leading to their exclusion from this consideration. Speech recognition was performed using Whisper large \cite{openai-whisper}, a transformer-based speech recognizer with a strong language model. Speech recognition was evaluated based on the percentage of correct answers per character

\subsection{Result}


The results are shown in Fig. \ref{fig:software_vs_hardware_wer_result}; the results for normal and whispered speech are shown on the left and right sides, respectively. For natural speech recorded with 80 dB background noise, the recognition accuracy for WhisperMask was over 20\% higher than that for pin microphone and earbuds. Notably, under the same noise condition, the recognition rate for WhisperMask without denoiser was higher by over 20\% than that for the microphones with denoiser.
The recognition accuracy for whispered speech recorded in an environment with 80 dB background noise was higher for WhisperMask by over 30\% than that for pin microphone and earbuds. After denoiser application, the recognition rate for whispered speech decreased by approximately 20\% in the 30, 40, and 60 dB environments. This is a significant decrease compared with that observed for natural speech, indicating that the denoiser is not well-suited for whispered speech.
Moreover, the recognition accuracy for whispered speech recorded in an environment with 80 dB background noise was higher by over 30\% for WhisperMask without denoiser compared with that for pin microphone and earbuds, suggesting the superiority of WhisperMask for recognizing whispered speech in noisy environments.

\section{DISCUSSION}
\subsection{Mechanism of noise reduction}
The difficulty of vibration may be one of the reasons behind the enhanced ability of WhisperMask to input speech in noisy environments. As shown in Figs. 6–8, despite the 80 dB sound coming from the dummy head speaker and the noise coming from the speaker, the maximum input on the microphone side was approximately 40 dB, indicating that a loud sound input is required. This phenomenon is likely because normal human speech is produced at approximately 60–80 dB, but when a sound source is closer to a microphone, louder voices are picked up. In fact, when measured at a distance of about 3 cm using a sound level meter, a normal voice becomes louder by approximately 80–90 dB. One of the key factors contributing to noise reduction is that voices are captured more loudly at a close proximity.

\subsection{Issues for daily use}
\subsubsection{Reusability} Microphones are devices intended for daily use and must be designed for durability. In this study, we asked nine users to use the proposed device; they were instructed to put on and take off their masks for each input session. We observed no performance issues related to mask usage. Moreover, the users wore the device over a woven mask, meaning the device can operate without direct contact with the mouth. Consequently, the proposed device was associated with fewer concerns in relation to contamination and hygiene compared with devices that are attached directly to the face. Furthermore, the vibrating component of the microphone can be detached from the circuitry and may be embedded in a protective material, making it washable without causing any issues.

\subsubsection{Noise when walking} We did not evaluate the impact of walking and other movements on the performance of WhisperMask. Motion artifacts may possibly introduce noise and affect the device’s performance. Further investigation is needed to assess the impact of user movements and develop strategies to avoid any associated noise.

\subsubsection{Blowing wind} iven that voice enters the microphone through air transmission, strong winds or turbulent airflow between the mouth and the microphone can disrupt voice recording. This issue also affects the performance of ordinary pin microphones, and a windshield may be necessary. Although we have not evaluated the effects of strong winds in this study, we have confirmed that sound can be collected even when the sensor is placed inside a non-woven fabric mask, which could provide some protection against wind.

In the future, we intend to further evaluate the impact of user movements, such as walking, and environmental factors, such as wind, on the performance of WhisperMask.


\section{Conclusion}



We propose WhisperMask, a mask-type electret condenser microphone that can clearly capture a user’s voice even in noisy environments compared with conventional microphones. We demonstrated WhisperMask’s acoustic characteristics by measuring its impulse response using swept-sine signals. Furthermore, we evaluated WhisperMask based on three key metrics: SNR, quality of recorded voices, and speech recognition rate. Across these metrics, WhisperMask significantly outperformed the conventional noise reduction methods, which involve either hardware- or software-based approaches.

The recognition rate for whispered speech recorded in an environment with 80 dB background noise was notably higher by over 30\% for WhisperMask than that for pin microphones and earbuds. Moreover, while a denoiser software decreased by approximately 20\% the other microphones’ recognition rate for whispered speech recorded with 30–60 dB background noise, WhisperMask maintained a high performance even without denoising, surpassing the performance of the other microphones by a large margin. These results highlight WhisperMask’s overwhelming superiority in capturing whispered speech under noisy conditions.

In conclusion, WhisperMask represents a significant advancement in the wearable microphone technology. By effectively addressing the challenge of capturing clear voice input, especially whispered speech, in high-noise environments, WhisperMask opens new possibilities for enhanced communication and interaction across a wide range of voice-based applications while preserving user privacy. Its lightweight mask-type form factor and exceptional noise suppression capabilities make it a promising tool for various real-world scenarios requiring reliable voice inputs.

\begin{acks}
This work was supported by JST ACT-X Grant JPMJAX23KG, JST Moonshot R\&D Grant JPMJMS2012, JST CREST Grant JPMJCR17A3, and the commissioned research by NICT Japan Grant JPJ012368C02901.
\end{acks}

\bibliographystyle{ACM-Reference-Format}
\bibliography{main}


\begin{thebibliography}{87}


\ifx \showCODEN    \undefined \def \showCODEN     #1{\unskip}     \fi
\ifx \showDOI      \undefined \def \showDOI       #1{#1}\fi
\ifx \showISBNx    \undefined \def \showISBNx     #1{\unskip}     \fi
\ifx \showISBNxiii \undefined \def \showISBNxiii  #1{\unskip}     \fi
\ifx \showISSN     \undefined \def \showISSN      #1{\unskip}     \fi
\ifx \showLCCN     \undefined \def \showLCCN      #1{\unskip}     \fi
\ifx \shownote     \undefined \def \shownote      #1{#1}          \fi
\ifx \showarticletitle \undefined \def \showarticletitle #1{#1}   \fi
\ifx \showURL      \undefined \def \showURL       {\relax}        \fi
\providecommand\bibfield[2]{#2}
\providecommand\bibinfo[2]{#2}
\providecommand\natexlab[1]{#1}
\providecommand\showeprint[2][]{arXiv:#2}

\bibitem[app({[n.\,d.]})]%
        {appleinsider.com}
 \bibinfo{year}{[n.\,d.]}\natexlab{}.
\newblock \showarticletitle{https://appleinsider.com/articles/21/03/30/apple-airpods-beats-dominated-audio-wearable-market-in-2020}.
\newblock


\bibitem[1534-3(2014)]%
        {mushra-ITU-R}
\bibfield{author}{\bibinfo{person}{B.~Series. Recommendation ITU-R~BS. 1534-3}.} \bibinfo{year}{2014}\natexlab{}.
\newblock \showarticletitle{method for the subjective assessment of intermediate quality level of audio systems}. In \bibinfo{booktitle}{\emph{International Telecommunication Union Radio Communication Assembly}}.
\newblock


\bibitem[Aubreville et~al\mbox{.}(2018)]%
        {mushra-ex1-denoise}
\bibfield{author}{\bibinfo{person}{M. Aubreville}, \bibinfo{person}{K. Ehrensperger}, \bibinfo{person}{A. Maier}, \bibinfo{person}{T. Rosenkranz}, \bibinfo{person}{B. Graf}, {and} \bibinfo{person}{H. Puder}.} \bibinfo{year}{2018}\natexlab{}.
\newblock \showarticletitle{Deep Denoising for Hearing Aid Applications}. In \bibinfo{booktitle}{\emph{2018 16th International Workshop on Acoustic Signal Enhancement (IWAENC)}}. \bibinfo{pages}{361--365}.
\newblock
\urldef\tempurl%
\url{https://doi.org/10.1109/IWAENC.2018.8521369}
\showDOI{\tempurl}


\bibitem[Bauer(1962)]%
        {A-century-of-microphones}
\bibfield{author}{\bibinfo{person}{Benjamin~B Bauer}.} \bibinfo{year}{1962}\natexlab{}.
\newblock \showarticletitle{A century of microphones}.
\newblock \bibinfo{journal}{\emph{Proceedings of the IRE}} \bibinfo{volume}{50}, \bibinfo{number}{5} (\bibinfo{year}{1962}), \bibinfo{pages}{719--729}.
\newblock


\bibitem[Beach et~al\mbox{.}(2019)]%
        {SleepMask}
\bibfield{author}{\bibinfo{person}{Christopher Beach}, \bibinfo{person}{Nazmul Karim}, {and} \bibinfo{person}{Alexander~J. Casson}.} \bibinfo{year}{2019}\natexlab{}.
\newblock \showarticletitle{A Graphene-Based Sleep Mask for Comfortable Wearable Eye Tracking}. In \bibinfo{booktitle}{\emph{2019 41st Annual International Conference of the IEEE Engineering in Medicine and Biology Society (EMBC)}}. \bibinfo{pages}{6693--6696}.
\newblock
\urldef\tempurl%
\url{https://doi.org/10.1109/EMBC.2019.8857198}
\showDOI{\tempurl}


\bibitem[Berkun and Cohen(2015)]%
        {snr_reverbing}
\bibfield{author}{\bibinfo{person}{Reuven Berkun} {and} \bibinfo{person}{Israel Cohen}.} \bibinfo{year}{2015}\natexlab{}.
\newblock \showarticletitle{Microphone array power ratio for quality assessment of reverberated speech}.
\newblock \bibinfo{journal}{\emph{EURASIP Journal on Advances in Signal Processing}} \bibinfo{volume}{2015}, \bibinfo{number}{1} (\bibinfo{year}{2015}), \bibinfo{pages}{49}.
\newblock
\showISBNx{1687-6180}
\urldef\tempurl%
\url{https://doi.org/10.1186/s13634-015-0233-y}
\showDOI{\tempurl}


\bibitem[Breithaupt et~al\mbox{.}(2008)]%
        {snr_novel}
\bibfield{author}{\bibinfo{person}{Colin Breithaupt}, \bibinfo{person}{Timo Gerkmann}, {and} \bibinfo{person}{Rainer Martin}.} \bibinfo{year}{2008}\natexlab{}.
\newblock \showarticletitle{A novel a priori SNR estimation approach based on selective cepstro-temporal smoothing}. In \bibinfo{booktitle}{\emph{2008 IEEE International Conference on Acoustics, Speech and Signal Processing}}. \bibinfo{pages}{4897--4900}.
\newblock
\urldef\tempurl%
\url{https://doi.org/10.1109/ICASSP.2008.4518755}
\showDOI{\tempurl}


\bibitem[Brumberg et~al\mbox{.}(2010)]%
        {EEG2_BCI}
\bibfield{author}{\bibinfo{person}{Jonathan~S. Brumberg}, \bibinfo{person}{Alfonso Nieto-Castanon}, \bibinfo{person}{Philip~R. Kennedy}, {and} \bibinfo{person}{Frank~H. Guenther}.} \bibinfo{year}{2010}\natexlab{}.
\newblock \showarticletitle{Brain–computer interfaces for speech communication}.
\newblock \bibinfo{journal}{\emph{Speech Communication}} \bibinfo{volume}{52}, \bibinfo{number}{4} (\bibinfo{year}{2010}), \bibinfo{pages}{367--379}.
\newblock
\showISSN{0167-6393}
\urldef\tempurl%
\url{https://doi.org/10.1016/j.specom.2010.01.001}
\showDOI{\tempurl}
\newblock
\shownote{Silent Speech Interfaces}.


\bibitem[Chatterjee et~al\mbox{.}(2022)]%
        {clearbuds}
\bibfield{author}{\bibinfo{person}{Ishan Chatterjee}, \bibinfo{person}{Maruchi Kim}, \bibinfo{person}{Vivek Jayaram}, \bibinfo{person}{Shyamnath Gollakota}, \bibinfo{person}{Ira Kemelmacher}, \bibinfo{person}{Shwetak Patel}, {and} \bibinfo{person}{Steven~M. Seitz}.} \bibinfo{year}{2022}\natexlab{}.
\newblock \showarticletitle{ClearBuds: Wireless Binaural Earbuds for Learning-Based Speech Enhancement}. In \bibinfo{booktitle}{\emph{Proceedings of the 20th Annual International Conference on Mobile Systems, Applications and Services}} (Portland, Oregon) \emph{(\bibinfo{series}{MobiSys '22})}. \bibinfo{publisher}{Association for Computing Machinery}, \bibinfo{address}{New York, NY, USA}, \bibinfo{pages}{384–396}.
\newblock
\showISBNx{9781450391856}
\urldef\tempurl%
\url{https://doi.org/10.1145/3498361.3538933}
\showDOI{\tempurl}


\bibitem[Choi et~al\mbox{.}(2005)]%
        {bss_review1}
\bibfield{author}{\bibinfo{person}{Seungjin Choi}, \bibinfo{person}{Andrzej Cichocki}, \bibinfo{person}{Hyung-Min Park}, {and} \bibinfo{person}{Soo-Young Lee}.} \bibinfo{year}{2005}\natexlab{}.
\newblock \showarticletitle{Blind source separation and independent component analysis: A review}.
\newblock \bibinfo{journal}{\emph{Neural Information Processing-Letters and Reviews}} \bibinfo{volume}{6}, \bibinfo{number}{1} (\bibinfo{year}{2005}), \bibinfo{pages}{1--57}.
\newblock


\bibitem[Cichocki and Phan(2009)]%
        {fastnmf}
\bibfield{author}{\bibinfo{person}{Andrzej Cichocki} {and} \bibinfo{person}{Anh-Huy Phan}.} \bibinfo{year}{2009}\natexlab{}.
\newblock \showarticletitle{Fast local algorithms for large scale nonnegative matrix and tensor factorizations}.
\newblock \bibinfo{journal}{\emph{IEICE transactions on fundamentals of electronics, communications and computer sciences}} \bibinfo{volume}{92}, \bibinfo{number}{3} (\bibinfo{year}{2009}), \bibinfo{pages}{708--721}.
\newblock


\bibitem[Cichocki et~al\mbox{.}(2006)]%
        {NMF}
\bibfield{author}{\bibinfo{person}{Andrzej Cichocki}, \bibinfo{person}{Rafal Zdunek}, {and} \bibinfo{person}{Shun-ichi Amari}.} \bibinfo{year}{2006}\natexlab{}.
\newblock \showarticletitle{New algorithms for non-negative matrix factorization in applications to blind source separation}. In \bibinfo{booktitle}{\emph{2006 IEEE International Conference on Acoustics Speech and Signal Processing Proceedings}}, Vol.~\bibinfo{volume}{5}. IEEE, \bibinfo{pages}{V--V}.
\newblock


\bibitem[Cohen(1962)]%
        {cohen1962statistical}
\bibfield{author}{\bibinfo{person}{J. Cohen}.} \bibinfo{year}{1962}\natexlab{}.
\newblock \showarticletitle{The statistical power of abnormal-social psychological research: A review}.
\newblock \bibinfo{journal}{\emph{The Journal of Abnormal and Social Psychology}} \bibinfo{volume}{65}, \bibinfo{number}{3} (\bibinfo{year}{1962}), \bibinfo{pages}{145--153}.
\newblock
\urldef\tempurl%
\url{https://doi.org/10.1037/h0045186}
\showDOI{\tempurl}


\bibitem[Corey and Singer(2018)]%
        {bss_microphone_work_well_with_numbers}
\bibfield{author}{\bibinfo{person}{Ryan~M Corey} {and} \bibinfo{person}{Andrew~C Singer}.} \bibinfo{year}{2018}\natexlab{}.
\newblock \showarticletitle{Speech separation using partially asynchronous microphone arrays without resampling}. In \bibinfo{booktitle}{\emph{2018 16th International Workshop on Acoustic Signal Enhancement (IWAENC)}}. IEEE, \bibinfo{pages}{1--9}.
\newblock


\bibitem[Csapó et~al\mbox{.}(2017)]%
        {DNN-Based-Ultrasound-to-Speech}
\bibfield{author}{\bibinfo{person}{Tamás Csapó}, \bibinfo{person}{Tamás Grósz}, \bibinfo{person}{Gábor Gosztolya}, \bibinfo{person}{László Tóth}, {and} \bibinfo{person}{Alexandra Markó}.} \bibinfo{year}{2017}\natexlab{}.
\newblock \showarticletitle{DNN-Based Ultrasound-to-Speech Conversion for a Silent Speech Interface}. In \bibinfo{booktitle}{\emph{interspeech 2017}}. \bibinfo{pages}{3672--3676}.
\newblock
\urldef\tempurl%
\url{https://doi.org/10.21437/Interspeech.2017-939}
\showDOI{\tempurl}


\bibitem[Denby et~al\mbox{.}(2010)]%
        {silentspeech_denby}
\bibfield{author}{\bibinfo{person}{B. Denby}, \bibinfo{person}{T. Schultz}, \bibinfo{person}{K. Honda}, \bibinfo{person}{T. Hueber}, \bibinfo{person}{J.M. Gilbert}, {and} \bibinfo{person}{J.S. Brumberg}.} \bibinfo{year}{2010}\natexlab{}.
\newblock \showarticletitle{Silent speech interfaces}.
\newblock \bibinfo{journal}{\emph{Speech Communication}} \bibinfo{volume}{52}, \bibinfo{number}{4} (\bibinfo{year}{2010}), \bibinfo{pages}{270 -- 287}.
\newblock
\showISSN{0167-6393}
\urldef\tempurl%
\url{https://doi.org/10.1016/j.specom.2009.08.002}
\showDOI{\tempurl}
\newblock
\shownote{Silent Speech Interfaces}.


\bibitem[Diener et~al\mbox{.}(2020)]%
        {emg_array}
\bibfield{author}{\bibinfo{person}{Lorenz Diener}, \bibinfo{person}{Mehrdad Roustay~Vishkasougheh}, {and} \bibinfo{person}{Tanja Schultz}.} \bibinfo{year}{2020}\natexlab{}.
\newblock \showarticletitle{{CSL-EMG\_Array: An Open Access Corpus for EMG-to-Speech Conversion}}. In \bibinfo{booktitle}{\emph{{INTERSPEECH} 2020 -- 21st Annual Conference of the International Speech Communication Association}}.
\newblock


\bibitem[Défossez et~al\mbox{.}(2020)]%
        {denoiser}
\bibfield{author}{\bibinfo{person}{Alexandre Défossez}, \bibinfo{person}{Gabriel Synnaeve}, {and} \bibinfo{person}{Yossi Adi}.} \bibinfo{year}{2020}\natexlab{}.
\newblock \showarticletitle{{Real Time Speech Enhancement in the Waveform Domain}}. In \bibinfo{booktitle}{\emph{Proc. Interspeech 2020}}. \bibinfo{pages}{3291--3295}.
\newblock
\urldef\tempurl%
\url{https://doi.org/10.21437/Interspeech.2020-2409}
\showDOI{\tempurl}


\bibitem[Electronics(2024)]%
        {LGPuriCare2024}
\bibfield{author}{\bibinfo{person}{LG Electronics}.} \bibinfo{year}{2024}\natexlab{}.
\newblock \bibinfo{title}{LG PuriCare™ Wearable Air Purifier (w/ VoiceON™) | LG Philippines}.
\newblock \bibinfo{howpublished}{\url{https://www.lg.com/ph/air-care/lg-ap551awfa}}.
\newblock
\newblock
\shownote{Accessed: 2024-02-10}.


\bibitem[Elenius(1980)]%
        {octaveband-smoothing}
\bibfield{author}{\bibinfo{person}{K Elenius}.} \bibinfo{year}{1980}\natexlab{}.
\newblock \bibinfo{title}{Long time average spectrum using a 1/3 octave filter bank}.
\newblock
\newblock


\bibitem[farina(2000)]%
        {swept-sine}
\bibfield{author}{\bibinfo{person}{angelo farina}.} \bibinfo{year}{2000}\natexlab{}.
\newblock \showarticletitle{simultaneous measurement of impulse response and distortion with a swept-sine technique}.
\newblock \bibinfo{journal}{\emph{journal of the audio engineering society}} (\bibinfo{date}{february} \bibinfo{year}{2000}).
\newblock


\bibitem[Fonseca et~al\mbox{.}(2018)]%
        {waveformer-dataset-1}
\bibfield{author}{\bibinfo{person}{Eduardo Fonseca}, \bibinfo{person}{Manoj Plakal}, \bibinfo{person}{Frederic Font}, \bibinfo{person}{Daniel P.~W. Ellis}, \bibinfo{person}{Xavier Favory}, \bibinfo{person}{Jordi Pons}, {and} \bibinfo{person}{Xavier Serra}.} \bibinfo{year}{2018}\natexlab{}.
\newblock \showarticletitle{General-purpose Tagging of Freesound Audio with AudioSet Labels: Task Description, Dataset, and Baseline}. In \bibinfo{booktitle}{\emph{Proceedings of the Detection and Classification of Acoustic Scenes and Events 2018 Workshop (DCASE2018)}}. \bibinfo{pages}{69--73}.
\newblock
\urldef\tempurl%
\url{https://arxiv.org/abs/1807.09902}
\showURL{%
\tempurl}


\bibitem[Freitas et~al\mbox{.}(2016)]%
        {An_Introduction_to_Silent_Speech_Interfaces}
\bibfield{author}{\bibinfo{person}{João Freitas}, \bibinfo{person}{António Teixeira}, \bibinfo{person}{Miguel Dias}, {and} \bibinfo{person}{Samuel Silva}.} \bibinfo{year}{2016}\natexlab{}.
\newblock \bibinfo{booktitle}{\emph{An Introduction to Silent Speech Interfaces}}.
\newblock
\showISBNx{978-3-319-40173-7}


\bibitem[Fukumoto(2018)]%
        {fukumoto}
\bibfield{author}{\bibinfo{person}{Masaaki Fukumoto}.} \bibinfo{year}{2018}\natexlab{}.
\newblock \showarticletitle{SilentVoice: Unnoticeable Voice Input by Ingressive Speech}. In \bibinfo{booktitle}{\emph{Proceedings of the 31st Annual ACM Symposium on User Interface Software and Technology}} (Berlin, Germany) \emph{(\bibinfo{series}{UIST '18})}. \bibinfo{publisher}{Association for Computing Machinery}, \bibinfo{address}{New York, NY, USA}, \bibinfo{pages}{237–246}.
\newblock
\showISBNx{9781450359481}
\urldef\tempurl%
\url{https://doi.org/10.1145/3242587.3242603}
\showDOI{\tempurl}


\bibitem[Gong(1995)]%
        {speech_recognition_survey}
\bibfield{author}{\bibinfo{person}{Yifan Gong}.} \bibinfo{year}{1995}\natexlab{}.
\newblock \showarticletitle{Speech recognition in noisy environments: A survey}.
\newblock \bibinfo{journal}{\emph{Speech Communication}} \bibinfo{volume}{16}, \bibinfo{number}{3} (\bibinfo{year}{1995}), \bibinfo{pages}{261--291}.
\newblock
\showISSN{0167-6393}
\urldef\tempurl%
\url{https://doi.org/10.1016/0167-6393(94)00059-J}
\showDOI{\tempurl}


\bibitem[Guo and Liang(2023)]%
        {TexonMask}
\bibfield{author}{\bibinfo{person}{Zengrong Guo} {and} \bibinfo{person}{Rong-Hao Liang}.} \bibinfo{year}{2023}\natexlab{}.
\newblock \showarticletitle{TexonMask: Facial Expression Recognition Using Textile Electrodes on Commodity Facemasks}. In \bibinfo{booktitle}{\emph{Proceedings of the 2023 CHI Conference on Human Factors in Computing Systems}} (Hamburg, Germany) \emph{(\bibinfo{series}{CHI '23})}. \bibinfo{publisher}{Association for Computing Machinery}, \bibinfo{address}{New York, NY, USA}, Article \bibinfo{articleno}{627}, \bibinfo{numpages}{15}~pages.
\newblock
\showISBNx{9781450394215}
\urldef\tempurl%
\url{https://doi.org/10.1145/3544548.3581295}
\showDOI{\tempurl}


\bibitem[Heittola et~al\mbox{.}(2019)]%
        {waveformer-dataset-2}
\bibfield{author}{\bibinfo{person}{Toni Heittola}, \bibinfo{person}{Annamaria Mesaros}, {and} \bibinfo{person}{Tuomas Virtanen}.} \bibinfo{year}{2019}\natexlab{}.
\newblock \bibinfo{booktitle}{\emph{{TAU Urban Acoustic Scenes 2019, Development dataset}}}.
\newblock
\urldef\tempurl%
\url{https://doi.org/10.5281/zenodo.2589280}
\showDOI{\tempurl}


\bibitem[Hirahara et~al\mbox{.}(2010)]%
        {NAM2}
\bibfield{author}{\bibinfo{person}{Tatsuya Hirahara}, \bibinfo{person}{Makoto Otani}, \bibinfo{person}{Shota Shimizu}, \bibinfo{person}{Tomoki Toda}, \bibinfo{person}{Keigo Nakamura}, \bibinfo{person}{Yoshitaka Nakajima}, {and} \bibinfo{person}{Kiyohiro Shikano}.} \bibinfo{year}{2010}\natexlab{}.
\newblock \showarticletitle{Silent-speech enhancement using body-conducted vocal-tract resonance signals}.
\newblock \bibinfo{journal}{\emph{Speech Communication}} \bibinfo{volume}{52}, \bibinfo{number}{4} (\bibinfo{year}{2010}), \bibinfo{pages}{301--313}.
\newblock
\showISSN{0167-6393}
\urldef\tempurl%
\url{https://doi.org/10.1016/j.specom.2009.12.001}
\showDOI{\tempurl}
\newblock
\shownote{Silent Speech Interfaces}.


\bibitem[Hiraki and Rekimoto(2021)]%
        {SilentMask}
\bibfield{author}{\bibinfo{person}{Hirotaka Hiraki} {and} \bibinfo{person}{Jun Rekimoto}.} \bibinfo{year}{2021}\natexlab{}.
\newblock \showarticletitle{SilentMask: Mask-Type Silent Speech Interface with Measurement of Mouth Movement}. In \bibinfo{booktitle}{\emph{Augmented Humans Conference 2021}} (Rovaniemi, Finland) \emph{(\bibinfo{series}{AHs'21})}. \bibinfo{publisher}{Association for Computing Machinery}, \bibinfo{address}{New York, NY, USA}, \bibinfo{pages}{86–90}.
\newblock
\showISBNx{9781450384285}
\urldef\tempurl%
\url{https://doi.org/10.1145/3458709.3458985}
\showDOI{\tempurl}


\bibitem[Hueber et~al\mbox{.}(2007)]%
        {ultrasound1}
\bibfield{author}{\bibinfo{person}{T. Hueber}, \bibinfo{person}{G. Aversano}, \bibinfo{person}{G. Chollet}, \bibinfo{person}{B. Denby}, \bibinfo{person}{G. Dreyfus}, \bibinfo{person}{Y. Oussar}, \bibinfo{person}{P. Roussel}, {and} \bibinfo{person}{M. Stone}.} \bibinfo{year}{2007}\natexlab{}.
\newblock \showarticletitle{Eigentongue Feature Extraction for an Ultrasound-Based Silent Speech Interface}. In \bibinfo{booktitle}{\emph{2007 IEEE International Conference on Acoustics, Speech and Signal Processing - ICASSP '07}}, Vol.~\bibinfo{volume}{1}. \bibinfo{pages}{I--1245--I--1248}.
\newblock
\urldef\tempurl%
\url{https://doi.org/10.1109/ICASSP.2007.366140}
\showDOI{\tempurl}


\bibitem[Inc.(2021)]%
        {Hushme}
\bibfield{author}{\bibinfo{person}{Hushme Inc.}} \bibinfo{year}{2021}\natexlab{}.
\newblock \bibinfo{title}{Hushme - The World's First Voice Mask for Smartphones}.
\newblock \bibinfo{howpublished}{\url{https://gethushme.com/}}.
\newblock
\newblock
\shownote{Accessed: 2024-02-10}.


\bibitem[Inc.(2023)]%
        {mutalk}
\bibfield{author}{\bibinfo{person}{Shiftall Inc.}} \bibinfo{year}{2023}\natexlab{}.
\newblock \bibinfo{title}{mutalk - Leakage voice suppression microphone}.
\newblock \bibinfo{howpublished}{\url{https://en.shiftall.net/products/mutalk}}.
\newblock
\newblock
\shownote{Accessed: 2024-02-10}.


\bibitem[Ingalls(1987)]%
        {Throat-microphone}
\bibfield{author}{\bibinfo{person}{Robert Ingalls}.} \bibinfo{year}{1987}\natexlab{}.
\newblock \showarticletitle{{Throat microphone}}.
\newblock \bibinfo{journal}{\emph{The Journal of the Acoustical Society of America}} \bibinfo{volume}{81}, \bibinfo{number}{3} (\bibinfo{date}{03} \bibinfo{year}{1987}), \bibinfo{pages}{809--809}.
\newblock
\showISSN{0001-4966}
\urldef\tempurl%
\url{https://doi.org/10.1121/1.394659}
\showDOI{\tempurl}
\showeprint{https://pubs.aip.org/asa/jasa/article-pdf/81/3/809/12095011/809\_1\_online.pdf}


\bibitem[Kapur et~al\mbox{.}(2018)]%
        {AlterEgo_emg4}
\bibfield{author}{\bibinfo{person}{Arnav Kapur}, \bibinfo{person}{Shreyas Kapur}, {and} \bibinfo{person}{Pattie Maes}.} \bibinfo{year}{2018}\natexlab{}.
\newblock \showarticletitle{AlterEgo: A Personalized Wearable Silent Speech Interface}. In \bibinfo{booktitle}{\emph{23rd International Conference on Intelligent User Interfaces}} (Tokyo, Japan) \emph{(\bibinfo{series}{IUI '18})}. \bibinfo{publisher}{Association for Computing Machinery}, \bibinfo{address}{New York, NY, USA}, \bibinfo{pages}{43–53}.
\newblock
\showISBNx{9781450349451}
\urldef\tempurl%
\url{https://doi.org/10.1145/3172944.3172977}
\showDOI{\tempurl}


\bibitem[Kawaguchi and Matsumoto(2022)]%
        {throatmic2}
\bibfield{author}{\bibinfo{person}{Junki Kawaguchi} {and} \bibinfo{person}{Mitsuharu Matsumoto}.} \bibinfo{year}{2022}\natexlab{}.
\newblock \showarticletitle{Noise Reduction Combining a General Microphone and a Throat Microphone}.
\newblock \bibinfo{journal}{\emph{Sensors}} \bibinfo{volume}{22}, \bibinfo{number}{12} (\bibinfo{year}{2022}).
\newblock
\showISSN{1424-8220}
\urldef\tempurl%
\url{https://doi.org/10.3390/s2212 s4473}
\showDOI{\tempurl}


\bibitem[Khanna et~al\mbox{.}(2021)]%
        {jawsense}
\bibfield{author}{\bibinfo{person}{Prerna Khanna}, \bibinfo{person}{Tanmay Srivastava}, \bibinfo{person}{Shijia Pan}, \bibinfo{person}{Shubham Jain}, {and} \bibinfo{person}{Phuc Nguyen}.} \bibinfo{year}{2021}\natexlab{}.
\newblock \showarticletitle{JawSense: Recognizing Unvoiced Sound Using a Low-Cost Ear-Worn System}. In \bibinfo{booktitle}{\emph{Proceedings of the 22nd International Workshop on Mobile Computing Systems and Applications}} (Virtual, United Kingdom) \emph{(\bibinfo{series}{HotMobile '21})}. \bibinfo{publisher}{Association for Computing Machinery}, \bibinfo{address}{New York, NY, USA}, \bibinfo{pages}{44–49}.
\newblock
\showISBNx{9781450383233}
\urldef\tempurl%
\url{https://doi.org/10.1145/3446382.3448363}
\showDOI{\tempurl}


\bibitem[Kim et~al\mbox{.}(2006)]%
        {IVA_Independent-vector-analysis}
\bibfield{author}{\bibinfo{person}{Taesu Kim}, \bibinfo{person}{Torbj{\o}rn Eltoft}, {and} \bibinfo{person}{Te-Won Lee}.} \bibinfo{year}{2006}\natexlab{}.
\newblock \showarticletitle{Independent vector analysis: An extension of ICA to multivariate components}. In \bibinfo{booktitle}{\emph{International conference on independent component analysis and signal separation}}. Springer, \bibinfo{pages}{165--172}.
\newblock


\bibitem[Kimura et~al\mbox{.}(2021)]%
        {silentspeller}
\bibfield{author}{\bibinfo{person}{Naoki Kimura}, \bibinfo{person}{Tan Gemicioglu}, \bibinfo{person}{Jonathan Womack}, \bibinfo{person}{Richard Li}, \bibinfo{person}{Yuhui Zhao}, \bibinfo{person}{Abdelkareem Bedri}, \bibinfo{person}{Alex Olwal}, \bibinfo{person}{Jun Rekimoto}, {and} \bibinfo{person}{Thad Starner}.} \bibinfo{year}{2021}\natexlab{}.
\newblock \bibinfo{booktitle}{\emph{Mobile, Hands-Free, Silent Speech Texting Using SilentSpeller}}.
\newblock \bibinfo{publisher}{Association for Computing Machinery}, \bibinfo{address}{New York, NY, USA}.
\newblock
\showISBNx{9781450380959}
\urldef\tempurl%
\url{https://doi.org/10.1145/3411763.3451552}
\showURL{%
\tempurl}


\bibitem[Kimura et~al\mbox{.}(2019)]%
        {sottovoce}
\bibfield{author}{\bibinfo{person}{Naoki Kimura}, \bibinfo{person}{Michinari Kono}, {and} \bibinfo{person}{Jun Rekimoto}.} \bibinfo{year}{2019}\natexlab{}.
\newblock \showarticletitle{SottoVoce: An Ultrasound Imaging-Based Silent Speech Interaction Using Deep Neural Networks}. In \bibinfo{booktitle}{\emph{Proceedings of the 2019 CHI Conference on Human Factors in Computing Systems}} (Glasgow, Scotland Uk) \emph{(\bibinfo{series}{CHI '19})}. \bibinfo{publisher}{Association for Computing Machinery}, \bibinfo{address}{New York, NY, USA}, \bibinfo{pages}{1–11}.
\newblock
\showISBNx{9781450359702}
\urldef\tempurl%
\url{https://doi.org/10.1145/3290605.3300376}
\showDOI{\tempurl}


\bibitem[Kitamura et~al\mbox{.}(2015)]%
        {irlma_kitamura2015efficient}
\bibfield{author}{\bibinfo{person}{Daichi Kitamura}, \bibinfo{person}{Nobutaka Ono}, \bibinfo{person}{Hiroshi Sawada}, \bibinfo{person}{Hirokazu Kameoka}, {and} \bibinfo{person}{Hiroshi Saruwatari}.} \bibinfo{year}{2015}\natexlab{}.
\newblock \showarticletitle{Efficient multichannel nonnegative matrix factorization exploiting rank-1 spatial model}. In \bibinfo{booktitle}{\emph{2015 IEEE International Conference on Acoustics, Speech and Signal Processing (ICASSP)}}. IEEE, \bibinfo{pages}{276--280}.
\newblock


\bibitem[Koizumi et~al\mbox{.}(2018)]%
        {koisumi_enhancement}
\bibfield{author}{\bibinfo{person}{Yuma Koizumi}, \bibinfo{person}{Kenta Niwa}, \bibinfo{person}{Yusuke Hioka}, \bibinfo{person}{Kazunori Kobayashi}, {and} \bibinfo{person}{Yoichi Haneda}.} \bibinfo{year}{2018}\natexlab{}.
\newblock \showarticletitle{DNN-Based Source Enhancement to Increase Objective Sound Quality Assessment Score}.
\newblock \bibinfo{journal}{\emph{IEEE/ACM Transactions on Audio, Speech, and Language Processing}} \bibinfo{volume}{26}, \bibinfo{number}{10} (\bibinfo{year}{2018}), \bibinfo{pages}{1780--1792}.
\newblock
\urldef\tempurl%
\url{https://doi.org/10.1109/TASLP.2018.2842156}
\showDOI{\tempurl}


\bibitem[Kumazaki and Inoue(2020)]%
        {TransEmotion}
\bibfield{author}{\bibinfo{person}{Ryoga Kumazaki} {and} \bibinfo{person}{Akifumi Inoue}.} \bibinfo{year}{2020}\natexlab{}.
\newblock \showarticletitle{Development and Evaluation of a Mask-Type Display Transforming the Wearer's Impression}. In \bibinfo{booktitle}{\emph{Proceedings of 31st Australian Conference on Human-Computer-Interaction}} (Fremantle, WA, Australia) \emph{(\bibinfo{series}{OzCHI '19})}. \bibinfo{publisher}{Association for Computing Machinery}, \bibinfo{address}{New York, NY, USA}, \bibinfo{pages}{568–571}.
\newblock
\showISBNx{9781450376969}
\urldef\tempurl%
\url{https://doi.org/10.1145/3369457.3369533}
\showDOI{\tempurl}


\bibitem[Kunimi et~al\mbox{.}(2022)]%
        {E-mask}
\bibfield{author}{\bibinfo{person}{Yusuke Kunimi}, \bibinfo{person}{Masa Ogata}, \bibinfo{person}{Hirotaka Hiraki}, \bibinfo{person}{Motoshi Itagaki}, \bibinfo{person}{Shusuke Kanazawa}, {and} \bibinfo{person}{Masaaki Mochimaru}.} \bibinfo{year}{2022}\natexlab{}.
\newblock \showarticletitle{E-MASK: A Mask-Shaped Interface for Silent Speech Interaction with Flexible Strain Sensors}. In \bibinfo{booktitle}{\emph{Augmented Humans 2022}} (Kashiwa, Chiba, Japan) \emph{(\bibinfo{series}{AHs 2022})}. \bibinfo{publisher}{Association for Computing Machinery}, \bibinfo{address}{New York, NY, USA}, \bibinfo{pages}{26–34}.
\newblock
\showISBNx{9781450396325}
\urldef\tempurl%
\url{https://doi.org/10.1145/3519391.3519399}
\showDOI{\tempurl}


\bibitem[Kusabuka and Indo(2020)]%
        {ibuki}
\bibfield{author}{\bibinfo{person}{Takahiro Kusabuka} {and} \bibinfo{person}{Takuya Indo}.} \bibinfo{year}{2020}\natexlab{}.
\newblock \showarticletitle{IBUKI: Gesture Input Method Based on Breathing}. In \bibinfo{booktitle}{\emph{Adjunct Publication of the 33rd Annual ACM Symposium on User Interface Software and Technology}} (Virtual Event, USA) \emph{(\bibinfo{series}{UIST '20 Adjunct})}. \bibinfo{publisher}{Association for Computing Machinery}, \bibinfo{address}{New York, NY, USA}, \bibinfo{pages}{102–104}.
\newblock
\showISBNx{9781450375153}
\urldef\tempurl%
\url{https://doi.org/10.1145/3379350.3416134}
\showDOI{\tempurl}


\bibitem[Lee et~al\mbox{.}(2020)]%
        {mascreen}
\bibfield{author}{\bibinfo{person}{Hyein Lee}, \bibinfo{person}{Yoonji Kim}, {and} \bibinfo{person}{Andrea Bianchi}.} \bibinfo{year}{2020}\natexlab{}.
\newblock \showarticletitle{MAScreen: Augmenting Speech with Visual Cues of Lip Motions, Facial Expressions, and Text Using a Wearable Display}. In \bibinfo{booktitle}{\emph{SIGGRAPH Asia 2020 Emerging Technologies}} (Virtual Event, Republic of Korea) \emph{(\bibinfo{series}{SA '20})}. \bibinfo{publisher}{Association for Computing Machinery}, \bibinfo{address}{New York, NY, USA}, Article \bibinfo{articleno}{2}, \bibinfo{numpages}{2}~pages.
\newblock
\showISBNx{9781450381109}
\urldef\tempurl%
\url{https://doi.org/10.1145/3415255.3422886}
\showDOI{\tempurl}


\bibitem[Li et~al\mbox{.}(2019)]%
        {TongueBoard}
\bibfield{author}{\bibinfo{person}{Richard Li}, \bibinfo{person}{Jason Wu}, {and} \bibinfo{person}{Thad Starner}.} \bibinfo{year}{2019}\natexlab{}.
\newblock \showarticletitle{TongueBoard: An Oral Interface for Subtle Input}. In \bibinfo{booktitle}{\emph{Proceedings of the 10th Augmented Human International Conference 2019}} (Reims, France) \emph{(\bibinfo{series}{AH2019})}. \bibinfo{publisher}{Association for Computing Machinery}, \bibinfo{address}{New York, NY, USA}, Article \bibinfo{articleno}{1}, \bibinfo{numpages}{9}~pages.
\newblock
\showISBNx{9781450365475}
\urldef\tempurl%
\url{https://doi.org/10.1145/3311823.3311831}
\showDOI{\tempurl}


\bibitem[Liaqat et~al\mbox{.}(2021)]%
        {smartwatch_mic1}
\bibfield{author}{\bibinfo{person}{Daniyal Liaqat}, \bibinfo{person}{Salaar Liaqat}, \bibinfo{person}{Jun~Lin Chen}, \bibinfo{person}{Tina Sedaghat}, \bibinfo{person}{Moshe Gabel}, \bibinfo{person}{Frank Rudzicz}, {and} \bibinfo{person}{Eyal de Lara}.} \bibinfo{year}{2021}\natexlab{}.
\newblock \showarticletitle{Coughwatch: Real-World Cough Detection using Smartwatches}. In \bibinfo{booktitle}{\emph{ICASSP 2021 - 2021 IEEE International Conference on Acoustics, Speech and Signal Processing (ICASSP)}}. \bibinfo{pages}{8333--8337}.
\newblock
\urldef\tempurl%
\url{https://doi.org/10.1109/ICASSP39728.2021.9414881}
\showDOI{\tempurl}


\bibitem[Liaqat et~al\mbox{.}(2018)]%
        {smartwatch_mic2}
\bibfield{author}{\bibinfo{person}{Daniyal Liaqat}, \bibinfo{person}{Robert Wu}, \bibinfo{person}{Andrea Gershon}, \bibinfo{person}{Hisham Alshaer}, \bibinfo{person}{Frank Rudzicz}, {and} \bibinfo{person}{Eyal de Lara}.} \bibinfo{year}{2018}\natexlab{}.
\newblock \showarticletitle{Challenges with Real-World Smartwatch Based Audio Monitoring}. In \bibinfo{booktitle}{\emph{Proceedings of the 4th ACM Workshop on Wearable Systems and Applications}} (Munich, Germany) \emph{(\bibinfo{series}{WearSys '18})}. \bibinfo{publisher}{Association for Computing Machinery}, \bibinfo{address}{New York, NY, USA}, \bibinfo{pages}{54–59}.
\newblock
\showISBNx{9781450358422}
\urldef\tempurl%
\url{https://doi.org/10.1145/3211960.3211977}
\showDOI{\tempurl}


\bibitem[MacKenzie and Soukoreff(2003)]%
        {MacKenzie_Soukoreff}
\bibfield{author}{\bibinfo{person}{I.~Scott MacKenzie} {and} \bibinfo{person}{R.~William Soukoreff}.} \bibinfo{year}{2003}\natexlab{}.
\newblock \showarticletitle{Phrase Sets for Evaluating Text Entry Techniques}. In \bibinfo{booktitle}{\emph{CHI '03 Extended Abstracts on Human Factors in Computing Systems}} (Ft. Lauderdale, Florida, USA) \emph{(\bibinfo{series}{CHI EA '03})}. \bibinfo{publisher}{Association for Computing Machinery}, \bibinfo{address}{New York, NY, USA}, \bibinfo{pages}{754–755}.
\newblock
\showISBNx{1581136374}
\urldef\tempurl%
\url{https://doi.org/10.1145/765891.765971}
\showDOI{\tempurl}


\bibitem[Makishima et~al\mbox{.}(2019)]%
        {idlma_makishima2019independent}
\bibfield{author}{\bibinfo{person}{Naoki Makishima}, \bibinfo{person}{Shinichi Mogami}, \bibinfo{person}{Norihiro Takamune}, \bibinfo{person}{Daichi Kitamura}, \bibinfo{person}{Hayato Sumino}, \bibinfo{person}{Shinnosuke Takamichi}, \bibinfo{person}{Hiroshi Saruwatari}, {and} \bibinfo{person}{Nobutaka Ono}.} \bibinfo{year}{2019}\natexlab{}.
\newblock \showarticletitle{Independent deeply learned matrix analysis for determined audio source separation}.
\newblock \bibinfo{journal}{\emph{IEEE/ACM Transactions on Audio, Speech, and Language Processing}} \bibinfo{volume}{27}, \bibinfo{number}{10} (\bibinfo{year}{2019}), \bibinfo{pages}{1601--1615}.
\newblock


\bibitem[Michelsanti et~al\mbox{.}(2021)]%
        {enhancement_review2}
\bibfield{author}{\bibinfo{person}{Daniel Michelsanti}, \bibinfo{person}{Zheng-Hua Tan}, \bibinfo{person}{Shi-Xiong Zhang}, \bibinfo{person}{Yong Xu}, \bibinfo{person}{Meng Yu}, \bibinfo{person}{Dong Yu}, {and} \bibinfo{person}{Jesper Jensen}.} \bibinfo{year}{2021}\natexlab{}.
\newblock \showarticletitle{An overview of deep-learning-based audio-visual speech enhancement and separation}.
\newblock \bibinfo{journal}{\emph{IEEE/ACM Transactions on Audio, Speech, and Language Processing}}  \bibinfo{volume}{29} (\bibinfo{year}{2021}), \bibinfo{pages}{1368--1396}.
\newblock


\bibitem[Nakajima et~al\mbox{.}(2003)]%
        {NAM0}
\bibfield{author}{\bibinfo{person}{Y. Nakajima}, \bibinfo{person}{H. Kashioka}, \bibinfo{person}{K. Shikano}, {and} \bibinfo{person}{N. Campbell}.} \bibinfo{year}{2003}\natexlab{}.
\newblock \showarticletitle{Non-audible murmur recognition input interface using stethoscopic microphone attached to the skin}. In \bibinfo{booktitle}{\emph{2003 IEEE International Conference on Acoustics, Speech, and Signal Processing, 2003. Proceedings. (ICASSP '03).}}, Vol.~\bibinfo{volume}{5}. \bibinfo{pages}{V--708}.
\newblock
\urldef\tempurl%
\url{https://doi.org/10.1109/ICASSP.2003.1200069}
\showDOI{\tempurl}


\bibitem[Nam et~al\mbox{.}(2020)]%
        {unmasked}
\bibfield{author}{\bibinfo{person}{Hye~Yeon Nam}, \bibinfo{person}{Iyleah Hernandez}, {and} \bibinfo{person}{Brendan Harmon}.} \bibinfo{year}{2020}\natexlab{}.
\newblock \showarticletitle{Unmasked}. In \bibinfo{booktitle}{\emph{Adjunct Publication of the 33rd Annual ACM Symposium on User Interface Software and Technology}} (Virtual Event, USA) \emph{(\bibinfo{series}{UIST '20 Adjunct})}. \bibinfo{publisher}{Association for Computing Machinery}, \bibinfo{address}{New York, NY, USA}, \bibinfo{pages}{111–113}.
\newblock
\showISBNx{9781450375153}
\urldef\tempurl%
\url{https://doi.org/10.1145/3379350.3416137}
\showDOI{\tempurl}


\bibitem[Pandey and Arif(2021)]%
        {liptype}
\bibfield{author}{\bibinfo{person}{Laxmi Pandey} {and} \bibinfo{person}{Ahmed~Sabbir Arif}.} \bibinfo{year}{2021}\natexlab{}.
\newblock \bibinfo{booktitle}{\emph{LipType: A Silent Speech Recognizer Augmented with an Independent Repair Model}}.
\newblock \bibinfo{publisher}{Association for Computing Machinery}, \bibinfo{address}{New York, NY, USA}.
\newblock
\showISBNx{9781450380966}
\urldef\tempurl%
\url{https://doi.org/10.1145/3411764.3445565}
\showURL{%
\tempurl}


\bibitem[Porbadnigk et~al\mbox{.}(2009)]%
        {EEG1}
\bibfield{author}{\bibinfo{person}{Anne Porbadnigk}, \bibinfo{person}{Marek Wester}, \bibinfo{person}{Jan-P Calliess}, {and} \bibinfo{person}{Tanja Schultz}.} \bibinfo{year}{2009}\natexlab{}.
\newblock \showarticletitle{EEG-based Speech Recognition - Impact of Temporal Effects.} \bibinfo{pages}{376--381}.
\newblock


\bibitem[Radford et~al\mbox{.}(2022)]%
        {openai-whisper}
\bibfield{author}{\bibinfo{person}{Alec Radford}, \bibinfo{person}{Jong~Wook Kim}, \bibinfo{person}{Tao Xu}, \bibinfo{person}{Greg Brockman}, \bibinfo{person}{Christine McLeavey}, {and} \bibinfo{person}{Ilya Sutskever}.} \bibinfo{year}{2022}\natexlab{}.
\newblock \bibinfo{title}{Robust Speech Recognition via Large-Scale Weak Supervision}.
\newblock
\newblock
\showeprint[arxiv]{2212.04356}~[eess.AS]


\bibitem[Reddy et~al\mbox{.}(2020)]%
        {denoiser_dataset-2}
\bibfield{author}{\bibinfo{person}{Chandan Reddy}, \bibinfo{person}{Ebrahim Beyrami}, \bibinfo{person}{Harishchandra Dubey}, \bibinfo{person}{Vishak Gopal}, \bibinfo{person}{Roger Cheng}, \bibinfo{person}{Ross Cutler}, \bibinfo{person}{Sergiy Matusevych}, \bibinfo{person}{Robert Aichner}, \bibinfo{person}{Ashkan Aazami}, \bibinfo{person}{Sebastian Braun}, \bibinfo{person}{Puneet Rana}, \bibinfo{person}{Sriram Srinivasan}, {and} \bibinfo{person}{Johannes Gehrke}.} \bibinfo{year}{2020}\natexlab{}.
\newblock \showarticletitle{The INTERSPEECH 2020 Deep Noise Suppression Challenge: Datasets, Subjective Speech Quality and Testing Framework}. In \bibinfo{booktitle}{\emph{Interspeech 2020}}.
\newblock


\bibitem[Rekimoto(2023)]%
        {mushra-ex3-wesper}
\bibfield{author}{\bibinfo{person}{Jun Rekimoto}.} \bibinfo{year}{2023}\natexlab{}.
\newblock \showarticletitle{WESPER: Zero-Shot and Realtime Whisper to Normal Voice Conversion for Whisper-Based Speech Interactions}. In \bibinfo{booktitle}{\emph{Proceedings of the 2023 CHI Conference on Human Factors in Computing Systems}} (Hamburg, Germany) \emph{(\bibinfo{series}{CHI '23})}. \bibinfo{publisher}{Association for Computing Machinery}, \bibinfo{address}{New York, NY, USA}, Article \bibinfo{articleno}{700}, \bibinfo{numpages}{12}~pages.
\newblock
\showISBNx{9781450394215}
\urldef\tempurl%
\url{https://doi.org/10.1145/3544548.3580706}
\showDOI{\tempurl}


\bibitem[Rekimoto and Nishimura(2021)]%
        {derma_ahs}
\bibfield{author}{\bibinfo{person}{Jun Rekimoto} {and} \bibinfo{person}{Yu Nishimura}.} \bibinfo{year}{2021}\natexlab{}.
\newblock \showarticletitle{Derma: Silent Speech Interaction Using Transcutaneous Motion Sensing}. In \bibinfo{booktitle}{\emph{Augmented Humans Conference 2021}} (Rovaniemi, Finland) \emph{(\bibinfo{series}{AHs'21})}. \bibinfo{publisher}{Association for Computing Machinery}, \bibinfo{address}{New York, NY, USA}, \bibinfo{pages}{91–100}.
\newblock
\showISBNx{9781450384285}
\urldef\tempurl%
\url{https://doi.org/10.1145/3458709.3458941}
\showDOI{\tempurl}


\bibitem[R\"{o}ddiger et~al\mbox{.}(2022)]%
        {earbuds_sensing}
\bibfield{author}{\bibinfo{person}{Tobias R\"{o}ddiger}, \bibinfo{person}{Christopher Clarke}, \bibinfo{person}{Paula Breitling}, \bibinfo{person}{Tim Schneegans}, \bibinfo{person}{Haibin Zhao}, \bibinfo{person}{Hans Gellersen}, {and} \bibinfo{person}{Michael Beigl}.} \bibinfo{year}{2022}\natexlab{}.
\newblock \showarticletitle{Sensing with Earables: A Systematic Literature Review and Taxonomy of Phenomena}.
\newblock  \bibinfo{volume}{6}, \bibinfo{number}{3}, Article \bibinfo{articleno}{135} (\bibinfo{date}{sep} \bibinfo{year}{2022}), \bibinfo{numpages}{57}~pages.
\newblock
\urldef\tempurl%
\url{https://doi.org/10.1145/3550314}
\showDOI{\tempurl}


\bibitem[R{\"o}ddiger et~al\mbox{.}(2022)]%
        {opensource_earbuds1}
\bibfield{author}{\bibinfo{person}{Tobias R{\"o}ddiger}, \bibinfo{person}{Tobias King}, \bibinfo{person}{Dylan~Ray Roodt}, \bibinfo{person}{Christopher Clarke}, {and} \bibinfo{person}{Michael Beigl}.} \bibinfo{year}{2022}\natexlab{}.
\newblock \showarticletitle{Openearable: Open hardware earable sensing platform}. In \bibinfo{booktitle}{\emph{Adjunct Proceedings of the 2022 ACM International Joint Conference on Pervasive and Ubiquitous Computing and the 2022 ACM International Symposium on Wearable Computers}}. \bibinfo{pages}{246--251}.
\newblock


\bibitem[Ronneberger et~al\mbox{.}(2015)]%
        {u-net}
\bibfield{author}{\bibinfo{person}{Olaf Ronneberger}, \bibinfo{person}{Philipp Fischer}, {and} \bibinfo{person}{Thomas Brox}.} \bibinfo{year}{2015}\natexlab{}.
\newblock \showarticletitle{U-Net: Convolutional Networks for Biomedical Image Segmentation}. In \bibinfo{booktitle}{\emph{Medical Image Computing and Computer-Assisted Intervention -- MICCAI 2015}}, \bibfield{editor}{\bibinfo{person}{Nassir Navab}, \bibinfo{person}{Joachim Hornegger}, \bibinfo{person}{William~M. Wells}, {and} \bibinfo{person}{Alejandro~F. Frangi}} (Eds.). \bibinfo{publisher}{Springer International Publishing}, \bibinfo{address}{Cham}, \bibinfo{pages}{234--241}.
\newblock
\showISBNx{978-3-319-24574-4}


\bibitem[Sakashita et~al\mbox{.}(2016)]%
        {yadori}
\bibfield{author}{\bibinfo{person}{Mose Sakashita}, \bibinfo{person}{Keisuke Kawahara}, \bibinfo{person}{Amy Koike}, \bibinfo{person}{Kenta Suzuki}, \bibinfo{person}{Ippei Suzuki}, {and} \bibinfo{person}{Yoichi Ochiai}.} \bibinfo{year}{2016}\natexlab{}.
\newblock \showarticletitle{Yadori: Mask-Type User Interface for Manipulation of Puppets}. In \bibinfo{booktitle}{\emph{ACM SIGGRAPH 2016 Emerging Technologies}} (Anaheim, California) \emph{(\bibinfo{series}{SIGGRAPH '16})}. \bibinfo{publisher}{Association for Computing Machinery}, \bibinfo{address}{New York, NY, USA}, Article \bibinfo{articleno}{23}, \bibinfo{numpages}{1}~pages.
\newblock
\showISBNx{9781450343725}
\urldef\tempurl%
\url{https://doi.org/10.1145/2929464.2929478}
\showDOI{\tempurl}


\bibitem[Sawada et~al\mbox{.}(2019)]%
        {bss_survey_sawada_ono_kameoka_kitamura_saruwatari_2019}
\bibfield{author}{\bibinfo{person}{Hiroshi Sawada}, \bibinfo{person}{Nobutaka Ono}, \bibinfo{person}{Hirokazu Kameoka}, \bibinfo{person}{Daichi Kitamura}, {and} \bibinfo{person}{Hiroshi Saruwatari}.} \bibinfo{year}{2019}\natexlab{}.
\newblock \showarticletitle{A review of blind source separation methods: two converging routes to ILRMA originating from ICA and NMF}.
\newblock \bibinfo{journal}{\emph{APSIPA Transactions on Signal and Information Processing}}  \bibinfo{volume}{8} (\bibinfo{year}{2019}), \bibinfo{pages}{e12}.
\newblock
\urldef\tempurl%
\url{https://doi.org/10.1017/ATSIP.2019.5}
\showDOI{\tempurl}


\bibitem[Schoeffler et~al\mbox{.}(2018)]%
        {webmushra}
\bibfield{author}{\bibinfo{person}{Michael Schoeffler}, \bibinfo{person}{Sarah Bartoschek}, \bibinfo{person}{Fabian-Robert Stöter}, \bibinfo{person}{Marlene Roess}, \bibinfo{person}{Susanne Westphal}, \bibinfo{person}{Bernd Edler}, {and} \bibinfo{person}{Jürgen Herre}.} \bibinfo{year}{2018}\natexlab{}.
\newblock \showarticletitle{webMUSHRA — A Comprehensive Framework for Web-based Listening Tests}.
\newblock \bibinfo{journal}{\emph{Journal of Open Research Software}} (\bibinfo{date}{Feb} \bibinfo{year}{2018}).
\newblock
\urldef\tempurl%
\url{https://doi.org/10.5334/jors.187}
\showDOI{\tempurl}


\bibitem[Schulte et~al\mbox{.}(2020)]%
        {ASR_in_operating_room}
\bibfield{author}{\bibinfo{person}{Antonia Schulte}, \bibinfo{person}{Rodrigo Suarez-Ibarrola}, \bibinfo{person}{Daniel Wegen}, \bibinfo{person}{Philippe-Fabian Pohlmann}, \bibinfo{person}{Elina Petersen}, {and} \bibinfo{person}{Arkadiusz Miernik}.} \bibinfo{year}{2020}\natexlab{}.
\newblock \showarticletitle{Automatic speech recognition in the operating room - An essential contemporary tool or a redundant gadget? A survey evaluation among physicians in form of a qualitative study}.
\newblock \bibinfo{journal}{\emph{Ann Med Surg (Lond)}}  \bibinfo{volume}{59} (\bibinfo{year}{2020}), \bibinfo{pages}{81--85}.
\newblock
\urldef\tempurl%
\url{https://doi.org/10.1016/j.amsu.2020.09.015}
\showDOI{\tempurl}


\bibitem[Shimizu et~al\mbox{.}(2009)]%
        {nam_microphone}
\bibfield{author}{\bibinfo{person}{Shota Shimizu}, \bibinfo{person}{Makoto Otani}, {and} \bibinfo{person}{Tatsuya Hirahara}.} \bibinfo{year}{2009}\natexlab{}.
\newblock \showarticletitle{Frequency characteristics of several non-audible murmur (NAM) microphones}.
\newblock \bibinfo{journal}{\emph{Acoustical Science and Technology}} \bibinfo{volume}{30}, \bibinfo{number}{2} (\bibinfo{year}{2009}), \bibinfo{pages}{139--142}.
\newblock
\urldef\tempurl%
\url{https://doi.org/10.1250/ast.30.139}
\showDOI{\tempurl}


\bibitem[Stoller et~al\mbox{.}(2018)]%
        {Wave-u-net}
\bibfield{author}{\bibinfo{person}{Daniel Stoller}, \bibinfo{person}{Sebastian Ewert}, {and} \bibinfo{person}{Simon Dixon}.} \bibinfo{year}{2018}\natexlab{}.
\newblock \showarticletitle{Wave-u-net: A multi-scale neural network for end-to-end audio source separation}.
\newblock \bibinfo{journal}{\emph{arXiv preprint arXiv:1806.03185}} (\bibinfo{year}{2018}).
\newblock


\bibitem[Su et~al\mbox{.}(2023)]%
        {liplearner}
\bibfield{author}{\bibinfo{person}{Zixiong Su}, \bibinfo{person}{Shitao Fang}, {and} \bibinfo{person}{Jun Rekimoto}.} \bibinfo{year}{2023}\natexlab{}.
\newblock \showarticletitle{LipLearner: Customizable Silent Speech Interactions on Mobile Devices}. In \bibinfo{booktitle}{\emph{Proceedings of the 2023 CHI Conference on Human Factors in Computing Systems}}. \bibinfo{pages}{1--21}.
\newblock


\bibitem[Subakan et~al\mbox{.}(2021)]%
        {sepformer}
\bibfield{author}{\bibinfo{person}{Cem Subakan}, \bibinfo{person}{Mirco Ravanelli}, \bibinfo{person}{Samuele Cornell}, \bibinfo{person}{Mirko Bronzi}, {and} \bibinfo{person}{Jianyuan Zhong}.} \bibinfo{year}{2021}\natexlab{}.
\newblock \showarticletitle{Attention Is All You Need In Speech Separation}. In \bibinfo{booktitle}{\emph{ICASSP 2021 - 2021 IEEE International Conference on Acoustics, Speech and Signal Processing (ICASSP)}}. \bibinfo{pages}{21--25}.
\newblock
\urldef\tempurl%
\url{https://doi.org/10.1109/ICASSP39728.2021.9413901}
\showDOI{\tempurl}


\bibitem[Sun et~al\mbox{.}(2018)]%
        {lip-interact}
\bibfield{author}{\bibinfo{person}{Ke Sun}, \bibinfo{person}{Chun Yu}, \bibinfo{person}{Weinan Shi}, \bibinfo{person}{Lan Liu}, {and} \bibinfo{person}{Yuanchun Shi}.} \bibinfo{year}{2018}\natexlab{}.
\newblock \showarticletitle{Lip-Interact: Improving Mobile Device Interaction with Silent Speech Commands}. In \bibinfo{booktitle}{\emph{Proceedings of the 31st Annual ACM Symposium on User Interface Software and Technology}} (Berlin, Germany) \emph{(\bibinfo{series}{UIST '18})}. \bibinfo{publisher}{Association for Computing Machinery}, \bibinfo{address}{New York, NY, USA}, \bibinfo{pages}{581–593}.
\newblock
\showISBNx{9781450359481}
\urldef\tempurl%
\url{https://doi.org/10.1145/3242587.3242599}
\showDOI{\tempurl}


\bibitem[Suzuki et~al\mbox{.}(2020)]%
        {mouthgesture}
\bibfield{author}{\bibinfo{person}{Yutaro Suzuki}, \bibinfo{person}{Kodai Sekimori}, \bibinfo{person}{Yuki Yamato}, \bibinfo{person}{Yusuke Yamasaki}, \bibinfo{person}{Buntarou Shizuki}, {and} \bibinfo{person}{Shin Takahashi}.} \bibinfo{year}{2020}\natexlab{}.
\newblock \showarticletitle{A Mouth Gesture Interface Featuring a Mutual-Capacitance Sensor Embedded in a Surgical Mask}. In \bibinfo{booktitle}{\emph{Human-Computer Interaction. Multimodal and Natural Interaction}}, \bibfield{editor}{\bibinfo{person}{Masaaki Kurosu}} (Ed.). \bibinfo{publisher}{Springer International Publishing}, \bibinfo{address}{Cham}, \bibinfo{pages}{154--165}.
\newblock
\showISBNx{978-3-030-49062-1}


\bibitem[Thibodeau et~al\mbox{.}(2021)]%
        {mask-is-difficult-to-speech-recognition1}
\bibfield{author}{\bibinfo{person}{Linda~M Thibodeau}, \bibinfo{person}{Rachel~B Thibodeau-Nielsen}, \bibinfo{person}{Chi Mai~Quynh Tran}, {and} \bibinfo{person}{Regina~Tangerino de Souza~Jacob}.} \bibinfo{year}{2021}\natexlab{}.
\newblock \showarticletitle{Communicating during COVID-19: The effect of transparent masks for speech recognition in noise}.
\newblock \bibinfo{journal}{\emph{Ear and Hearing}} \bibinfo{volume}{42}, \bibinfo{number}{4} (\bibinfo{year}{2021}), \bibinfo{pages}{772--781}.
\newblock


\bibitem[Tipparaju et~al\mbox{.}(2020a)]%
        {Respiration}
\bibfield{author}{\bibinfo{person}{Vishal~Varun Tipparaju}, \bibinfo{person}{Di Wang}, \bibinfo{person}{Jingjing Yu}, \bibinfo{person}{Fang Chen}, \bibinfo{person}{Francis Tsow}, \bibinfo{person}{Erica Forzani}, \bibinfo{person}{Nongjian Tao}, {and} \bibinfo{person}{Xiaojun Xian}.} \bibinfo{year}{2020}\natexlab{a}.
\newblock \showarticletitle{Respiration pattern recognition by wearable mask device}.
\newblock \bibinfo{journal}{\emph{Biosensors and Bioelectronics}}  \bibinfo{volume}{169} (\bibinfo{year}{2020}), \bibinfo{pages}{112590}.
\newblock
\showISSN{0956-5663}
\urldef\tempurl%
\url{https://doi.org/10.1016/j.bios.2020.112590}
\showDOI{\tempurl}


\bibitem[Tipparaju et~al\mbox{.}(2020b)]%
        {ReliableBreathing}
\bibfield{author}{\bibinfo{person}{Vishal~Varun Tipparaju}, \bibinfo{person}{Xiaojun Xian}, \bibinfo{person}{Devon Bridgeman}, \bibinfo{person}{Di Wang}, \bibinfo{person}{Francis Tsow}, \bibinfo{person}{Erica Forzani}, {and} \bibinfo{person}{Nongjian Tao}.} \bibinfo{year}{2020}\natexlab{b}.
\newblock \showarticletitle{Reliable Breathing Tracking With Wearable Mask Device}.
\newblock \bibinfo{journal}{\emph{IEEE Sensors Journal}} \bibinfo{volume}{20}, \bibinfo{number}{10} (\bibinfo{year}{2020}), \bibinfo{pages}{5510--5518}.
\newblock
\urldef\tempurl%
\url{https://doi.org/10.1109/JSEN.2020.2969635}
\showDOI{\tempurl}


\bibitem[Toscano and Toscano(2021)]%
        {mask-is-difficult-to-speech-recognition2}
\bibfield{author}{\bibinfo{person}{Joseph~C Toscano} {and} \bibinfo{person}{Cheyenne~M Toscano}.} \bibinfo{year}{2021}\natexlab{}.
\newblock \showarticletitle{Effects of face masks on speech recognition in multi-talker babble noise}.
\newblock \bibinfo{journal}{\emph{PloS one}} \bibinfo{volume}{16}, \bibinfo{number}{2} (\bibinfo{year}{2021}), \bibinfo{pages}{e0246842}.
\newblock


\bibitem[Valentini-Botinhao(2017)]%
        {denoiser_dataset-1}
\bibfield{author}{\bibinfo{person}{Cassia Valentini-Botinhao}.} \bibinfo{year}{2017}\natexlab{}.
\newblock \showarticletitle{Noisy speech database for training speech enhancement algorithms and TTS models}. In \bibinfo{booktitle}{\emph{University of Edinburgh. School of Informatics. Centre for Speech Technology Research (CSTR).}}
\newblock
\urldef\tempurl%
\url{https://doi.org/10.7488/ds/2117.}
\showDOI{\tempurl}


\bibitem[Veluri et~al\mbox{.}(2023)]%
        {waveformer}
\bibfield{author}{\bibinfo{person}{Bandhav Veluri}, \bibinfo{person}{Justin Chan}, \bibinfo{person}{Malek Itani}, \bibinfo{person}{Tuochao Chen}, \bibinfo{person}{Takuya Yoshioka}, {and} \bibinfo{person}{Shyamnath Gollakota}.} \bibinfo{year}{2023}\natexlab{}.
\newblock \showarticletitle{Real-Time Target Sound Extraction}. In \bibinfo{booktitle}{\emph{ICASSP 2023 - 2023 IEEE International Conference on Acoustics, Speech and Signal Processing (ICASSP)}}. \bibinfo{pages}{1--5}.
\newblock
\urldef\tempurl%
\url{https://doi.org/10.1109/ICASSP49357.2023.10094573}
\showDOI{\tempurl}


\bibitem[Vijayan et~al\mbox{.}(2017)]%
        {throatmic1}
\bibfield{author}{\bibinfo{person}{Amritha Vijayan}, \bibinfo{person}{Bipil~Mary Mathai}, \bibinfo{person}{Karthik Valsalan}, \bibinfo{person}{Riyanka~Raji Johnson}, \bibinfo{person}{Lani~Rachel Mathew}, {and} \bibinfo{person}{K. Gopakumar}.} \bibinfo{year}{2017}\natexlab{}.
\newblock \showarticletitle{Throat microphone speech recognition using mfcc}. In \bibinfo{booktitle}{\emph{2017 International Conference on Networks \& Advances in Computational Technologies (NetACT)}}. \bibinfo{pages}{392--395}.
\newblock
\urldef\tempurl%
\url{https://doi.org/10.1109/NETACT.2017.8076802}
\showDOI{\tempurl}


\bibitem[Wand. et~al\mbox{.}(2013)]%
        {EMG1_wand}
\bibfield{author}{\bibinfo{person}{Michael Wand.}, \bibinfo{person}{Christopher Schulte.}, \bibinfo{person}{Matthias Janke.}, {and} \bibinfo{person}{Tanja Schultz.}} \bibinfo{year}{2013}\natexlab{}.
\newblock \showarticletitle{Array-based Electromyographic Silent Speech Interface}. In \bibinfo{booktitle}{\emph{Proceedings of the International Conference on Bio-inspired Systems and Signal Processing - BIOSIGNALS, (BIOSTEC 2013)}}. INSTICC, \bibinfo{publisher}{SciTePress}, \bibinfo{pages}{89--96}.
\newblock
\showISBNx{978-989-8565-36-5}
\showISSN{2184-4305}
\urldef\tempurl%
\url{https://doi.org/10.5220/0004252400890096}
\showDOI{\tempurl}


\bibitem[Wand and Schultz(2011)]%
        {EMG0_wand}
\bibfield{author}{\bibinfo{person}{Michael Wand} {and} \bibinfo{person}{Tanja Schultz}.} \bibinfo{year}{2011}\natexlab{}.
\newblock \showarticletitle{Session-independent EMG-based Speech Recognition.}, In \bibinfo{booktitle}{Proceedings of Biosignals 2011}.
\newblock \bibinfo{journal}{\emph{BIOSIGNALS 2011 - Proceedings of the International Conference on Bio-Inspired Systems and Signal Processing}}, \bibinfo{pages}{295--300}.
\newblock


\bibitem[Yamamoto et~al\mbox{.}(2023)]%
        {masktrap}
\bibfield{author}{\bibinfo{person}{Takumi Yamamoto}, \bibinfo{person}{Katsutoshi Masai}, \bibinfo{person}{Anusha Withana}, {and} \bibinfo{person}{Yuta Sugiura}.} \bibinfo{year}{2023}\natexlab{}.
\newblock \showarticletitle{Masktrap: Designing and Identifying Gestures to Transform Mask Strap into an Input Interface}. In \bibinfo{booktitle}{\emph{Proceedings of the 28th International Conference on Intelligent User Interfaces}} (Sydney, NSW, Australia) \emph{(\bibinfo{series}{IUI '23})}. \bibinfo{publisher}{Association for Computing Machinery}, \bibinfo{address}{New York, NY, USA}, \bibinfo{pages}{762–775}.
\newblock
\showISBNx{9798400701061}
\urldef\tempurl%
\url{https://doi.org/10.1145/3581641.3584062}
\showDOI{\tempurl}


\bibitem[Yoshitaka et~al\mbox{.}(2005)]%
        {NAM1}
\bibfield{author}{\bibinfo{person}{NAKAJIMA Yoshitaka}, \bibinfo{person}{KASHIOKA Hideki}, \bibinfo{person}{CAMPBELL Nick}, {and} \bibinfo{person}{SHIKANO Kiyohiro}.} \bibinfo{year}{2005}\natexlab{}.
\newblock \showarticletitle{Non-Audible Murmur (NAM) Recognition}.
\newblock \bibinfo{journal}{\emph{IEICE TRANSACTIONS on Information and Systems}} \bibinfo{volume}{E89-D}, \bibinfo{number}{1} (\bibinfo{year}{2005}).
\newblock


\bibitem[Yuliani et~al\mbox{.}(2021)]%
        {enhancement_review1}
\bibfield{author}{\bibinfo{person}{Asri~Rizki Yuliani}, \bibinfo{person}{M~Faizal Amri}, \bibinfo{person}{Endang Suryawati}, \bibinfo{person}{Ade Ramdan}, {and} \bibinfo{person}{Hilman~Ferdinandus Pardede}.} \bibinfo{year}{2021}\natexlab{}.
\newblock \showarticletitle{Speech enhancement using deep learning methods: A review}.
\newblock \bibinfo{journal}{\emph{Jurnal Elektronika dan Telekomunikasi}} \bibinfo{volume}{21}, \bibinfo{number}{1} (\bibinfo{year}{2021}), \bibinfo{pages}{19--26}.
\newblock


\bibitem[Zalkow et~al\mbox{.}(2023)]%
        {mushra-ex2-tts}
\bibfield{author}{\bibinfo{person}{Frank Zalkow}, \bibinfo{person}{Prachi Govalkar}, \bibinfo{person}{Meinard Müller}, \bibinfo{person}{Emanuël A.~P. Habets}, {and} \bibinfo{person}{Christian Dittmar}.} \bibinfo{year}{2023}\natexlab{}.
\newblock \showarticletitle{Evaluating Speech–Phoneme Alignment and its Impact on Neural Text-To-Speech Synthesis}. In \bibinfo{booktitle}{\emph{ICASSP 2023 - 2023 IEEE International Conference on Acoustics, Speech and Signal Processing (ICASSP)}}. \bibinfo{pages}{1--5}.
\newblock
\urldef\tempurl%
\url{https://doi.org/10.1109/ICASSP49357.2023.10097248}
\showDOI{\tempurl}


\bibitem[Zhang et~al\mbox{.}(2017)]%
        {smartring_mic1}
\bibfield{author}{\bibinfo{person}{Cheng Zhang}, \bibinfo{person}{Anandghan Waghmare}, \bibinfo{person}{Pranav Kundra}, \bibinfo{person}{Yiming Pu}, \bibinfo{person}{Scott Gilliland}, \bibinfo{person}{Thomas Ploetz}, \bibinfo{person}{Thad~E. Starner}, \bibinfo{person}{Omer~T. Inan}, {and} \bibinfo{person}{Gregory~D. Abowd}.} \bibinfo{year}{2017}\natexlab{}.
\newblock \showarticletitle{FingerSound: Recognizing Unistroke Thumb Gestures Using a Ring}.
\newblock \bibinfo{journal}{\emph{Proc. ACM Interact. Mob. Wearable Ubiquitous Technol.}} \bibinfo{volume}{1}, \bibinfo{number}{3}, Article \bibinfo{articleno}{120} (\bibinfo{date}{sep} \bibinfo{year}{2017}), \bibinfo{numpages}{19}~pages.
\newblock
\urldef\tempurl%
\url{https://doi.org/10.1145/3130985}
\showDOI{\tempurl}


\bibitem[Zhang et~al\mbox{.}(2020)]%
        {endophasia}
\bibfield{author}{\bibinfo{person}{Yongzhao Zhang}, \bibinfo{person}{Wei-Hsiang Huang}, \bibinfo{person}{Chih-Yun Yang}, \bibinfo{person}{Wen-Ping Wang}, \bibinfo{person}{Yi-Chao Chen}, \bibinfo{person}{Chuang-Wen You}, \bibinfo{person}{Da-Yuan Huang}, \bibinfo{person}{Guangtao Xue}, {and} \bibinfo{person}{Jiadi Yu}.} \bibinfo{year}{2020}\natexlab{}.
\newblock \showarticletitle{Endophasia: Utilizing Acoustic-Based Imaging for Issuing Contact-Free Silent Speech Commands}.
\newblock \bibinfo{journal}{\emph{Proc. ACM Interact. Mob. Wearable Ubiquitous Technol.}} \bibinfo{volume}{4}, \bibinfo{number}{1}, Article \bibinfo{articleno}{37} (\bibinfo{date}{March} \bibinfo{year}{2020}), \bibinfo{numpages}{26}~pages.
\newblock
\urldef\tempurl%
\url{https://doi.org/10.1145/3381008}
\showDOI{\tempurl}


\end{thebibliography}

\end{document}